\documentclass[twocolumn]{aastex63}

\shorttitle{Direct Optimal Mapping}
\shortauthors{Xu et al.}

\usepackage{CJK}

\newcommand{\dg}{$^\circ$}
\newcommand{\mb}[1]{$\mathbf{#1}$}

\begin{document}
\begin{CJK*}{UTF8}{gbsn}

\title{Direct Optimal Mapping for 21\,cm Cosmology:\\ A Demonstration with the Hydrogen Epoch of Reionization Array}

\correspondingauthor{Zhilei Xu (徐智磊)}
\email{zhileixu@mit.edu\\zhileixu@space.mit.edu}

\author[0000-0001-5112-2567]{Zhilei  Xu (徐智磊)}
\affiliation{MIT Kavli Institute, Massachusetts Institute of Technology, Cambridge, MA}

\author[0000-0002-4117-570X]{Jacqueline N. Hewitt}
\affiliation{MIT Kavli Institute, Massachusetts Institute of Technology, Cambridge, MA}
\affiliation{Department of Physics, Massachusetts Institute of Technology, Cambridge, MA}

\author[0000-0002-3839-0230]{Kai-Feng  Chen}
\affiliation{Department of Physics, Massachusetts Institute of Technology, Cambridge, MA}

\author{Honggeun  Kim}
\affiliation{Department of Physics, Massachusetts Institute of Technology, Cambridge, MA}

\author[0000-0003-3336-9958]{Joshua S. Dillon}
\affiliation{Department of Astronomy, University of California, Berkeley, CA}

\author[0000-0002-8211-1892]{Nicholas S. Kern}
\affiliation{Department of Astronomy, University of California, Berkeley, CA}
\affiliation{Department of Physics, Massachusetts Institute of Technology, Cambridge, MA}

\author[0000-0001-7694-4030]{Miguel F. Morales}
\affiliation{Department of Physics, University of Washington, Seattle, WA}

\author[0000-0001-7532-645X]{Bryna J. Hazelton}
\affiliation{Department of Physics, University of Washington, Seattle, WA}
\affiliation{eScience Institute, University of Washington, Seattle, WA}

\author{Ruby  Byrne}
\affiliation{Department of Physics, University of Washington, Seattle, WA}

\author{Nicolas  Fagnoni}
\affiliation{Cavendish Astrophysics, University of Cambridge, Cambridge, UK}

\author{Eloy  de~Lera~Acedo}
\affiliation{Cavendish Astrophysics, University of Cambridge, Cambridge, UK}

\author{Zara  Abdurashidova}
\affiliation{Department of Astronomy, University of California, Berkeley, CA}

\author{Tyrone  Adams}
\affiliation{South African Radio Astronomy Observatory, Black River Park, 2 Fir Street, Observatory, Cape Town, 7925, South Africa}

\author[0000-0002-4810-666X]{James E. Aguirre}
\affiliation{Department of Physics and Astronomy, University of Pennsylvania, Philadelphia, PA}

\author{Paul  Alexander}
\affiliation{Cavendish Astrophysics, University of Cambridge, Cambridge, UK}

\author{Zaki S. Ali}
\affiliation{Department of Astronomy, University of California, Berkeley, CA}

\author{Rushelle  Baartman}
\affiliation{South African Radio Astronomy Observatory, Black River Park, 2 Fir Street, Observatory, Cape Town, 7925, South Africa}

\author{Yanga  Balfour}
\affiliation{South African Radio Astronomy Observatory, Black River Park, 2 Fir Street, Observatory, Cape Town, 7925, South Africa}

\author[0000-0001-9428-8233]{Adam P. Beardsley}
\affiliation{School of Earth and Space Exploration, Arizona State University, Tempe, AZ}
\affiliation{Department of Physics, Winona State University, Winona, MN}
\affiliation{NSF Astronomy and Astrophysics Postdoctoral Fellow}

\author[0000-0002-0916-7443]{Gianni  Bernardi}
\affiliation{INAF-Istituto di Radioastronomia, via Gobetti 101, 40129 Bologna, Italy}
\affiliation{Department of Physics and Electronics, Rhodes University, PO Box 94, Grahamstown, 6140, South Africa}
\affiliation{South African Radio Astronomy Observatory, Black River Park, 2 Fir Street, Observatory, Cape Town, 7925, South Africa}

\author{Tashalee S. Billings}
\affiliation{Department of Physics and Astronomy, University of Pennsylvania, Philadelphia, PA}

\author[0000-0002-8475-2036]{Judd D. Bowman}
\affiliation{School of Earth and Space Exploration, Arizona State University, Tempe, AZ}

\author{Richard F. Bradley}
\affiliation{National Radio Astronomy Observatory, Charlottesville, VA}

\author[0000-0001-5668-3101]{Philip  Bull}
\affiliation{Queen Mary University London, London E1 4NS, UK}
\affiliation{Department of Physics and Astronomy,  University of Western Cape, Cape Town, 7535, South Africa}

\author{Jacob  Burba}
\affiliation{Department of Physics, Brown University, Providence, RI}

\author{Steven  Carey}
\affiliation{Cavendish Astrophysics, University of Cambridge, Cambridge, UK}

\author[0000-0001-6647-3861]{Chris L. Carilli}
\affiliation{National Radio Astronomy Observatory, Socorro, NM 87801, USA}

\author{Carina  Cheng}
\affiliation{Department of Astronomy, University of California, Berkeley, CA}

\author[0000-0003-3197-2294]{David R. DeBoer}
\affiliation{Radio Astronomy Lab, University of California, Berkeley, CA}

\author{Matt  Dexter}
\affiliation{Radio Astronomy Lab, University of California, Berkeley, CA}

\author{Nico  Eksteen}
\affiliation{South African Radio Astronomy Observatory, Black River Park, 2 Fir Street, Observatory, Cape Town, 7925, South Africa}

\author{John  Ely}
\affiliation{Cavendish Astrophysics, University of Cambridge, Cambridge, UK}

\author[0000-0002-0086-7363]{Aaron  Ewall-Wice}
\affiliation{Department of Astronomy, University of California, Berkeley, CA}
\affiliation{Department of Physics, University of California, Berkeley, CA}

\author{Randall  Fritz}
\affiliation{South African Radio Astronomy Observatory, Black River Park, 2 Fir Street, Observatory, Cape Town, 7925, South Africa}

\author[0000-0002-0658-1243]{Steven R. Furlanetto}
\affiliation{Department of Physics and Astronomy, University of California, Los Angeles, CA}

\author{Kingsley  Gale-Sides}
\affiliation{Cavendish Astrophysics, University of Cambridge, Cambridge, UK}

\author{Brian  Glendenning}
\affiliation{National Radio Astronomy Observatory, Socorro, NM}

\author[0000-0002-0829-167X]{Deepthi  Gorthi}
\affiliation{Department of Astronomy, University of California, Berkeley, CA}

\author[0000-0002-4085-2094]{Bradley  Greig}
\affiliation{School of Physics, University of Melbourne, Parkville, VIC 3010, Australia}

\author{Jasper  Grobbelaar}
\affiliation{South African Radio Astronomy Observatory, Black River Park, 2 Fir Street, Observatory, Cape Town, 7925, South Africa}

\author{Ziyaad  Halday}
\affiliation{South African Radio Astronomy Observatory, Black River Park, 2 Fir Street, Observatory, Cape Town, 7925, South Africa}

\author{Jack  Hickish}
\affiliation{Radio Astronomy Lab, University of California, Berkeley, CA}

\author[0000-0002-0917-2269]{Daniel C. Jacobs}
\affiliation{School of Earth and Space Exploration, Arizona State University, Tempe, AZ}

\author{Austin  Julius}
\affiliation{South African Radio Astronomy Observatory, Black River Park, 2 Fir Street, Observatory, Cape Town, 7925, South Africa}

\author{MacCalvin  Kariseb}
\affiliation{South African Radio Astronomy Observatory, Black River Park, 2 Fir Street, Observatory, Cape Town, 7925, South Africa}

\author[0000-0002-1876-272X]{Joshua  Kerrigan}
\affiliation{Department of Physics, Brown University, Providence, RI}

\author[0000-0003-0953-313X]{Piyanat  Kittiwisit}
\affiliation{Department of Physics and Astronomy,  University of Western Cape, Cape Town, 7535, South Africa}

\author[0000-0001-6744-5328]{Saul A. Kohn}
\affiliation{Department of Physics and Astronomy, University of Pennsylvania, Philadelphia, PA}

\author[0000-0002-2950-2974]{Matthew  Kolopanis}
\affiliation{School of Earth and Space Exploration, Arizona State University, Tempe, AZ}

\author{Adam  Lanman}
\affiliation{Department of Physics, Brown University, Providence, RI}

\author[0000-0002-4693-0102]{Paul  La~Plante}
\affiliation{Department of Astronomy, University of California, Berkeley, CA}
\affiliation{Department of Physics and Astronomy, University of Pennsylvania, Philadelphia, PA}

\author[0000-0001-6876-0928]{Adrian  Liu}
\affiliation{Department of Astronomy, University of California, Berkeley, CA}
\affiliation{Department of Physics and McGill Space Institute, McGill University, 3600 University Street, Montreal, QC H3A 2T8, Canada}

\author{Anita  Loots}
\affiliation{South African Radio Astronomy Observatory, Black River Park, 2 Fir Street, Observatory, Cape Town, 7925, South Africa}

\author[0000-0001-8108-0986]{Yin-Zhe Ma}
\affiliation{School of Chemistry and Physics, University of KwaZulu-Natal, Westville Campus, Private Bag X54001, Durban 4000, South Africa}

\author{David Harold~Edward MacMahon}
\affiliation{Radio Astronomy Lab, University of California, Berkeley, CA}

\author{Lourence  Malan}
\affiliation{South African Radio Astronomy Observatory, Black River Park, 2 Fir Street, Observatory, Cape Town, 7925, South Africa}

\author{Cresshim  Malgas}
\affiliation{South African Radio Astronomy Observatory, Black River Park, 2 Fir Street, Observatory, Cape Town, 7925, South Africa}

\author{Keith  Malgas}
\affiliation{South African Radio Astronomy Observatory, Black River Park, 2 Fir Street, Observatory, Cape Town, 7925, South Africa}

\author{Bradley  Marero}
\affiliation{South African Radio Astronomy Observatory, Black River Park, 2 Fir Street, Observatory, Cape Town, 7925, South Africa}

\author{Zachary E. Martinot}
\affiliation{Department of Physics and Astronomy, University of Pennsylvania, Philadelphia, PA}

\author[0000-0003-3374-1772]{Andrei  Mesinger}
\affiliation{Scuola Normale Superiore, 56126 Pisa, PI, Italy}

\author{Mathakane  Molewa}
\affiliation{South African Radio Astronomy Observatory, Black River Park, 2 Fir Street, Observatory, Cape Town, 7925, South Africa}

\author{Tshegofalang  Mosiane}
\affiliation{South African Radio Astronomy Observatory, Black River Park, 2 Fir Street, Observatory, Cape Town, 7925, South Africa}

\author[0000-0003-3059-3823]{Steven G. Murray}
\affiliation{School of Earth and Space Exploration, Arizona State University, Tempe, AZ}

\author[0000-0001-7776-7240]{Abraham R. Neben}
\affiliation{Department of Physics, Massachusetts Institute of Technology, Cambridge, MA}

\author{Bojan  Nikolic}
\affiliation{Cavendish Astrophysics, University of Cambridge, Cambridge, UK}

\author{Hans  Nuwegeld}
\affiliation{South African Radio Astronomy Observatory, Black River Park, 2 Fir Street, Observatory, Cape Town, 7925, South Africa}

\author[0000-0002-5400-8097]{Aaron R. Parsons}
\affiliation{Department of Astronomy, University of California, Berkeley, CA}

\author[0000-0002-9457-1941]{Nipanjana  Patra}
\affiliation{Department of Astronomy, University of California, Berkeley, CA}

\author{Samantha  Pieterse}
\affiliation{South African Radio Astronomy Observatory, Black River Park, 2 Fir Street, Observatory, Cape Town, 7925, South Africa}

\author[0000-0002-3492-0433]{Jonathan C. Pober}
\affiliation{Department of Physics, Brown University, Providence, RI}

\author{Nima  Razavi-Ghods}
\affiliation{Cavendish Astrophysics, University of Cambridge, Cambridge, UK}

\author{James  Robnett}
\affiliation{National Radio Astronomy Observatory, Socorro, NM 87801, USA}

\author{Kathryn  Rosie}
\affiliation{South African Radio Astronomy Observatory, Black River Park, 2 Fir Street, Observatory, Cape Town, 7925, South Africa}

\author[0000-0002-2871-0413]{Peter  Sims}
\affiliation{Department of Physics and McGill Space Institute, McGill University, 3600 University Street, Montreal, QC H3A 2T8, Canada}

\author{Craig  Smith}
\affiliation{South African Radio Astronomy Observatory, Black River Park, 2 Fir Street, Observatory, Cape Town, 7925, South Africa}

\author{Hilton  Swarts}
\affiliation{South African Radio Astronomy Observatory, Black River Park, 2 Fir Street, Observatory, Cape Town, 7925, South Africa}

\author[0000-0003-1602-7868]{Nithyanandan  Thyagarajan}
\affiliation{Commonwealth Scientific and Industrial Research Organisation (CSIRO), Space \& Astronomy, P. O. Box 1130, Bentley, WA 6102, Australia}
\affiliation{National Radio Astronomy Observatory, Socorro, NM 87801, USA}

\author{Pieter  Van~Van~Wyngaarden}
\affiliation{South African Radio Astronomy Observatory, Black River Park, 2 Fir Street, Observatory, Cape Town, 7925, South Africa}

\author[0000-0003-3734-3587]{Peter K.~G. Williams}
\affiliation{Center for Astrophysics, Harvard \& Smithsonian, Cambridge, MA}
\affiliation{American Astronomical Society, Washington, DC}

\author{Haoxuan  Zheng}
\affiliation{Department of Physics, Massachusetts Institute of Technology, Cambridge, MA}

\collaboration{81}{(HERA Collaboration)}

\begin{abstract}
Motivated by the desire for wide-field images with well-defined statistical properties for 21\,cm cosmology, we implement an optimal mapping pipeline that 
computes a maximum likelihood estimator for the sky using the interferometric measurement equation.
We demonstrate this ``direct optimal mapping'' with data from the Hydrogen Epoch of Reionization (HERA) Phase I observations.
After validating the pipeline with simulated data, we develop a maximum likelihood figure-of-merit for comparing four sky models at 166\,MHz with a bandwidth of 100\,kHz.
The HERA data agree with the GLEAM catalogs~\citep{wayth/etal:2015} to $<$\,10\%.
After subtracting the GLEAM point sources, the HERA data discriminate between the different continuum sky models, providing most support for the model of \citet{byrne/etal:2021}. 
We report the computation cost for mapping the HERA Phase I data and project the computation for the HERA 320-antenna data; both are feasible with a modern server.
The algorithm is broadly applicable to other interferometers and is valid for wide-field and non-coplanar arrays.

\end{abstract}



\section{Introduction} \label{sec:intro}

Observations of the 21\,cm spectral line of neutral hydrogen during the epoch of reionization (EoR), cosmic dawn, and the Dark Ages have the potential to transform our understanding of the universe.
Goals of current experiments are to map the process of the formation
and evolution of the first stars, galaxies and black holes,  to further constrain
the prevailing $\Lambda$CDM cosmology~\citep{bennett/etal:1996, bennett/etal:2013, hinshaw/etal:2013, planck/etal:2020}, and to search for evidence of physics
beyond the $\Lambda$CDM paradigm. For reviews see \citep{liu/shaw:2020,mesinger:2016,morales/wyithe:2010,pritchard/loeb:2012}.
Current and recent interferometric
experiments aiming to detect cosmological 21\,cm signals include CHIME~\citep{bandura/etal:2014}, HIRAX~\citep{newburgh/etal:2016}, PAPER~\citep{parsons/etal:2010}, MWA~\citep{tingay/etal:2013}, LOFAR~\citep{vanhaarlem/etal:2013}, and HERA~\citep{deboer/etal:2017}. 

The ultimate goal of precision cosmology with the 21\,cm line is a quantitative comparison between theoretical predictions of neutral hydrogen structures and their measurements at radio wavelengths.
This confrontation between theory and experiments using interferometers
is beginning with two-point statistics (two-point correlation function or power spectrum), and is likely to develop further with higher-order statistics and, finally, direct evaluation of properties of 3D (two angular dimensions and one frequency dimension) image cubes.

Radio interferometers measure the coherence, or visibility, between signals received by pairs of antennas in an array.
For co-planar arrays and small fields of view, the relationship between the measured visibilities and the brightness distribution on the sky is  approximately a 2D Fourier transform~\citep*{thompson/moran/swenson:2017}, and is therefore closely
related to the power spectrum~\citep{morales/hewitt:2004}.
Wider fields  of view can be accommodated with the 2D Fourier technique by implementing corrections for neglecting the ``$w$-term'' in the interferometric phase
~\citep{cornwell/golap:2005,cornwell/voronkov/humphreys:2012,barry/etal:2019, ye/etal:2021}.
Current limits on the 21\,cm power spectrum have been derived by analyses that make use of the image-visibility Fourier relationship~\citep{dillon/etal:2014, dillon/etal:2015b, beardsley/etal:2016, trott/etal:2016, patil/etal:2017, barry/etal:2019, Li/etal:2019, rahimi/etal:2021} and by analyses that work directly with the visibility data through the delay spectrum approach~\citep{parsons/etal:2012, kolopnis/etal:2019, hera/etal:2021}.
The two approaches are complementary, essentially measuring different statistics in the sky~\citep{morales/etal:2019}.

Images mapped from visibilities, and convolved by the array synthesized beam, are called dirty images.
To deconvolve the dirty images, the CLEAN~\citep{hogbom:1974, clark:1980, cornwell:2008, rau/cornwell:2011} deconvolution algorithm is frequently used.  
The resulting CLEAN model is a list of deconvolved bright sources, which are the focus of astronomy and astrophysics. 
However, the focus of 21\,cm precision cosmology is the faint diffuse emission which is not corrected by CLEAN.
A deconvolution approach is required to treat all pixels equally across the image.
Even for bright sources, CLEAN is a heuristic, iterative process, and the statistical properties of the resulting image are not well known.
For cosmological studies, we need an algorithm that can map wide fields and that can reconstruct bright point sources and faint diffuse emission with equally well understood statistics.
Other ways to image and deconvolve images have been explored.
Some examples are Fast Holographic Deconvolution and other forward modeling techniques~\citep{sullivan/etal:2012, bernardi/etal:2013}
and $m$-mode mapping~\citep{shaw/etal:2014,eastwood/etal:2018}. For a recent review, see~\citet{liu/shaw:2020}.

In this paper we explore \textit{direct optimal mapping} (DOM) of interferometric data,
and we apply it to HERA data as a demonstration. 
By \textit{``direct''}, we mean that we do not make the assumptions that lead to the two-dimensional Fourier transform relationship between data and image, and
we take \textit{``optimal''} to mean that the mapping process does not lose information of model parameters. 
The optimal mapping approach was first explored in the context of CMB observations~\citep{tegmark/maps:1997}. 
It has more recently been extended to interferometric imaging for 21\,cm cosmology
\citep{morales/matejek:2009, sullivan/etal:2012, dillon/etal:2015,zheng/etal:2017b},
and considerations of optimal mapping are incorporated in the HERA design~\citep{dillon/parsons:2016}. 

Benefits of this approach include a data product in the image domain that potentially covers the full celestial sphere, 
full knowledge of the point spread function in all directions, and full knowledge of the covariance matrix relating map pixels.
Linear deconvolution, implemented through matrix operations, is in principle possible, treating point sources
and extended emission equally.  With the direct inversion of the instrument response, it is not necessary to grid
the data~\citep{barry/chokshi:2022}, it is not necessary to correct for neglect of the $w$-term, and the configuration of the antennas need not
be coplanar.

The paper is organized as follows.
We introduce the DOM general mathematical formalism in Section~\ref{sec:formalism}, and how we apply the formalism  to HERA data in Section~\ref{sec:application_to_hera}.
We validate the algorithm with simulated data in Section~\ref{sec:validation}.
In Section~\ref{sec:mapping_hera_data}, we use DOM to map HERA data, and evaluate four sky models.
In Section~\ref{sec:computation}, we assess the computational costs of DOM.
We conclude in Section~\ref{sec:conslusion}.

\section{Formalism} 
\label{sec:formalism}
Interferometers measure intensity as complex visibilities, defined as:
\begin{equation} \label{equ:visibility}
    V_{ij}(\nu) = \int B_{ij}(\mathbf{\hat s}) I (\mathbf{\hat s}) \exp \left( -i\, \frac{2\pi \nu}{c} \mathbf{b}_{ij} \cdot \mathbf{\hat s} \right) \, d\Omega,
\end{equation}
where $\mathbf{\hat s}$ is the unit vector pointing to a certain point on the sky, $B_{ij}(\mathbf{\hat s})$ is the product of two primary beams from the $i-j$ baseline\footnote{The primary beams are peak-normalized.}, 
$I(\mathbf{\hat s})$ is the specific intensity,
$\mathbf{b}_{ij}$ is the baseline vector, $\nu$ and $c$ are the observation frequency and the speed of light.

If we denote the sky as a map vector $\mathbf{m}$ and the measured visibilities to form a data vector $\mathbf{d}$, the mapping can be expressed by one matrix multiplication
\begin{equation} \label{equ:data_model}
    \mathbf{d} = \textbf{A} \mathbf{m} + \mathbf{n},
\end{equation}
where $\mathbf{A}$ is the measurement matrix and $\mathbf{n}$ is the noise vector for the visibilities.
The $\mathbf{A}$ matrix maps the sky to visibilities.

With this data model,
an optimal estimator of the sky is obtained with the following operation:
\begin{equation} \label{equ:optimal_mapping}
    \mathbf{\hat m} = \mathbf{D} \mathbf{A^\dagger} \mathbf{N^{-1}}  \mathbf{d},
\end{equation}
where $\mathbf{N} = \langle \mathbf{n} \mathbf{n}^\dagger \rangle$ is the noise covariance matrix, $\mathbf{D}$ in general transforms the raw map to the final estimation of the true sky, and  $\mathbf{\hat m}$  is an estimate of the true sky.
In this paper, we only format $\mathbf{D}$ as a normalization to physics units; we leave more complicated $\mathbf{D}$ formats, say including deconvolution, to future publications.
Even as a normalization matrix, $\mathbf{D}$ does not have a set form, and we discuss our choice of $\mathbf{D}$ in Section~\ref{subsec:map_normalization}.

DOM calculates direction-dependent PSFs for all pixels, expressed in a $N_\mathrm{pixel} \times N_\mathrm{pixel}$ matrix $\mathbf{P}$~\citep{dillon/etal:2015}:
\begin{equation} \label{equ:p_mat}
    \mathbf{P = D A^\dagger N^{-1} A}.
\end{equation}

For interferometers, pixel noise is highly correlated.
Therefore, the pixel covariance matrix is critical for
quantitative interpretation of the measured sky map.
One advantage from DOM is that it provides a pixel covariance matrix. 
The covariance matrix $\mathbf{C}$ is closely related to the PSF matrix~\citep{dillon/etal:2015}:
\begin{equation} \label{equ:covariance}
    \mathbf{C} = \mathbf{D} \mathbf{A^\dagger} \mathbf{N^{-1}}  \mathbf{A} \mathbf{D}^\mathrm{T} = \mathbf{P} \mathbf{D}^\mathrm{T}.
\end{equation}

\begin{figure*}
    \centering
    \includegraphics[width=\linewidth]{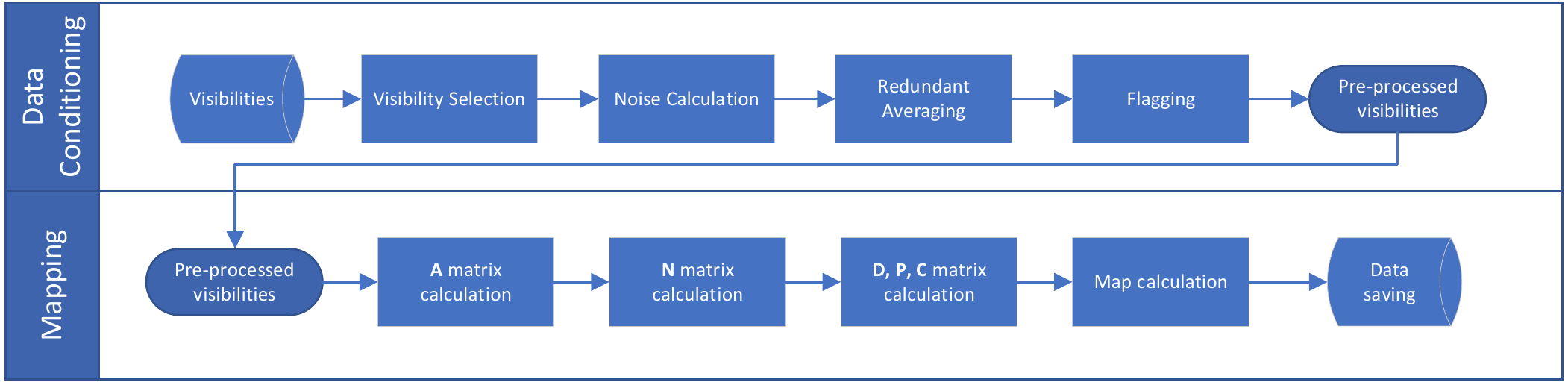}
    \caption{Diagram of the direct optimal mapping (DOM). The two major modules, \textit{data conditioning} and \textit{mapping}, are displayed on the left; the detailed steps are shown in the flowchart. More details of each step are described in Section~\ref{sec:application_to_hera}.}
    \label{fig:diagram}
\end{figure*}

To use DOM, we first need to construct the measurement matrix $\mathbf{A}$ and the noise matrix $\mathbf{N}$.
We divide the sky into HEALpixels~\citep{gorski/etal:2005} with the pixel size chosen to be much smaller than the synthesized beam. 
Then the integral in Equation~\ref{equ:visibility} becomes a summation:
\begin{equation} \label{equ:visibility_discrete}
    d_n = V_{n}(\nu) = \sum_k m_k B(\mathbf{\hat s}_{n, k}) \exp \left( -i\, \frac{2\pi \nu}{c} \mathbf{b}_{n} \mathbf{\hat s}_{n,k} \right),
\end{equation}
where the subscript $n$ represents all visibilities (folding $ij$ into one index), the subscript $k$ represents all pixels, $d_n$ is an element of the data vector \mb{d},
$\mathbf{\hat s}_{n,k}$ is the directional vector in the horizon coordinate system, and $m_k$ is the flux within one pixel calculated as $m_k = I_k\cdot \Delta \Omega$ ($\Delta \Omega$ is the solid angle of one HEALpixel). 
Please note that approximating spatially extended flux to the center of HEALpixels leads to errors, especially when the in-pixel flux distribution is strongly skewed by bright off-center point sources. 
This will be discussed later in the treatment of the GLEAM catalogs in Section~\ref{subssec:hera_map_vs_sky_models}.
In the above equation, we assume all the antenna pairs have the same primary beam $B$, and $\mathbf{\hat{s}}_{n, k}$ implicitly depends on time because of sky rotation.

Then the $\textbf{A}$ matrix is written element-wise:
\begin{eqnarray} \label{equ:a_matrix}
    A_{nk} 
    & =& B(\mathbf{\hat s}_{n, k}) \exp \left( -i \frac{2\pi\nu}{c} \mathbf{b}_n \cdot \mathbf{\hat s}_{n, k} \right)\\ 
    \label{equ:a_matrix_grouped}
    & =& B(\mathbf{\hat s}_{n, k}) \cdot \Phi_{n, k},
\end{eqnarray}
where the second line shows that the beam term $B(\mathbf{\hat s}_{n, k})$ and the fringe term $\Phi_{n, k}$ are separable.
We will use this feature in Section~\ref{subsec:map_normalization} for normalization.

\section{Application to HERA}
\label{sec:application_to_hera}

In this paper, we use HERA data from the Phase I observation season, with 39 operational antennas.
This is the dataset that was used to derive the Phase I EoR power spectrum limits, and we refer the reader to 
\cite{kern/etal:2019},
\cite{dillon/etal:2020}, \cite{kern/etal:2020}, 
\cite{kern/etal:2020b}, and \cite{hera/etal:2021} for a detailed discussion of the data preparation.
We only map the east-west polarization for this demonstration because \citet{byrne/etal:2021} shows that this patch of the sky is predominantly unpolarized.

Briefly, the dataset spans the dates of December 10th\,--\,28th, 2017, and the data were
calibrated, flagged, and binned by local sidereal time (LST).
Certain instrumental systematics, including crosstalk and cable reflection, were modeled and subtracted.
HERA is designed to have a redundant array
configuration~\citep{dillon/parsons:2016},
the resulting non-redundant baseline groups poorly sample the $uvw$ space, making HERA not an ideal instrument for imaging.
However, the focus of DOM is to make images with well understood
statistical properties, regardless of the resolving power of the images themselves.

We have developed a software package\footnote{The ``direct optimal mapping'' repository is developed on \href{https://github.com/HERA-Team/direct_optimal_mapping}{GitHub}; the first release version (v1.0.0, Zenodo, doi:\href{https://zenodo.org/badge/latestdoi/353426311
}{10.5281/zenodo.6984370}) is used for the analyses in this paper.} for the DOM algorithm.
The diagram in Figure~\ref{fig:diagram} illustrates steps of the algorithm.
Although the package is initially implemented in HERA data, it can be easily applied to other interferometric data.

\subsection{Calculating $\mathbf{A}$ and $\mathbf{N}$}
\label{subsec:matrices_calculation}

The sky pixels are defined in the \textit{equatorial coordinate system}, meaning that the pixels are independent of Earth rotation. 
For a zenith-pointing telescope, both the primary beam $B$ and the unit vectors $\mathbf{\hat s}_{n,k}$ are natively defined in the \textit{horizon coordinate system}.
At each time integration, the pixel locations are converted from the equatorial coordinate system to the horizon coordinate system (RA/Dec$\rightarrow$Az/El).
Then with the baseline vectors ($\mathbf{b}_n$ in Equation~\ref{equ:a_matrix}), all the elements of the $\mathbf{A}$ matrix are calculated.

The input data are first conditioned by selecting
visibilities, estimating the 
visibility noise, removing flagged data,
and averaging redundant baselines\footnote{Given the manageable amount of data, we do not perform redundant averaging for HERA Phase I data.
However, we will need redundant averaging for future HERA data with 320 antennas.}.
We estimate the noise for each visibility according to the radiometer equation~\citep{thompson/moran/swenson:2017, kern/etal:2020}
\begin{equation} \label{equ:noise_calc}
    \sigma_{n} = \frac{\sqrt{V_{ii} V_{jj}}}{\sqrt{\Delta \nu \Delta t}},
\end{equation}
where $\sigma_{n}$ is the estimate of the noise of the visibility between antenna $i$ and $j$, $V_{ii}$ and $V_{jj}$ are autocorrelations of antenna $i$ and $j$, $\Delta \nu$ is the bandwidth, and $\Delta t$ is the integration time.
When visibilities are redundant-averaged, the estimated noise for each redundant-averaged baseline is calculated as 
\begin{equation}
    \sigma_\mathrm{red.\ avg.} = \left( \sum_n \frac{1}{\sigma^2_n} \right)^{-1/2},
\end{equation}
where the summation runs over visibilities from redundant baselines within one redundant group.

The noise matrix $\mathbf{N}$ is constructed by filling the diagonal elements with the visibility noise squared, assuming off-diagonal elements are zero:
\begin{equation} \label{equ:noise_matrix}
    \mathbf{N} = \mathrm{diag}(\sigma_1^2, \sigma_1^2, \cdots, \sigma_{N_{vis}}^2),
\end{equation}
where $\sigma_n$ is the noise of the $n$th visibility.

\subsection{Map Normalization}
\label{subsec:map_normalization}

The direct optimal mapping formalism only requires the $\mathbf{D}$ matrix being non-singular to preserve all information of model parameters.
In practice, different normalization can be applied for different map-based analysis.

Without normalization, the mapping equation $\mathbf{\hat m = A^\dagger N^{-1} d}$ adds up contribution from all visibilities as a sum.
To calculate the average, we need the effective weights, which varies across pixels because of the primary beam.
In addition, sky drift complicates the weighting because it moves the primary beam on the sky.
Therefore, we need a weighting of the visibilities considering a moving primary beam for each sky pixel.

For HERA, the instrument observes the sky naturally with one primary beam  applied.
After that, noise is introduced in visibilities.
In direct optimal mapping, we apply another primary beam from multiplying $\mathbf{A^\dagger}$ in the mapping equation (Equation~\ref{equ:optimal_mapping}).
Therefore, the recovered sky has the primary beam applied \textit{twice}---one from observation and one from mapping.
However, the noise only has the primary beam applied \textit{once} from mapping.

Because of the difference, there does not exist an obvious way to correct for the primary beam effect in normalization.
We may correct for the primary beam \textit{twice}, and the recovered sky will have no primary beam applied; however, the noise will have one extra primary beam corrected, blowing up the noise far away from the beam center.
Alternatively, if we correct for the primary beam \textit{once}, the recovered sky will still have one primary beam applied, but the map noise will be free of primary beam effect.
Here we choose to correct for the primary beam \textit{once}, and define the normalization matrix $\mathbf{D}$ accordingly.

Inspired by \citet{barry/etal:2019}, we use the optimal mapping equation but replace the visibility data with a vector of ones for counting.
Then, we divide $\mathbf{\hat{m}}$ by the weight map for normalization.
However, the weight map has zero values because of the small-scale fringe term in the $\mathbf{A}$ matrix, which are numerically unstable in the denominator.
Instead, we use only the primary beam term in the $\mathbf{A}$ matrix to construct the weight map.

\begin{figure}
    \centering
    \includegraphics[width=\linewidth]{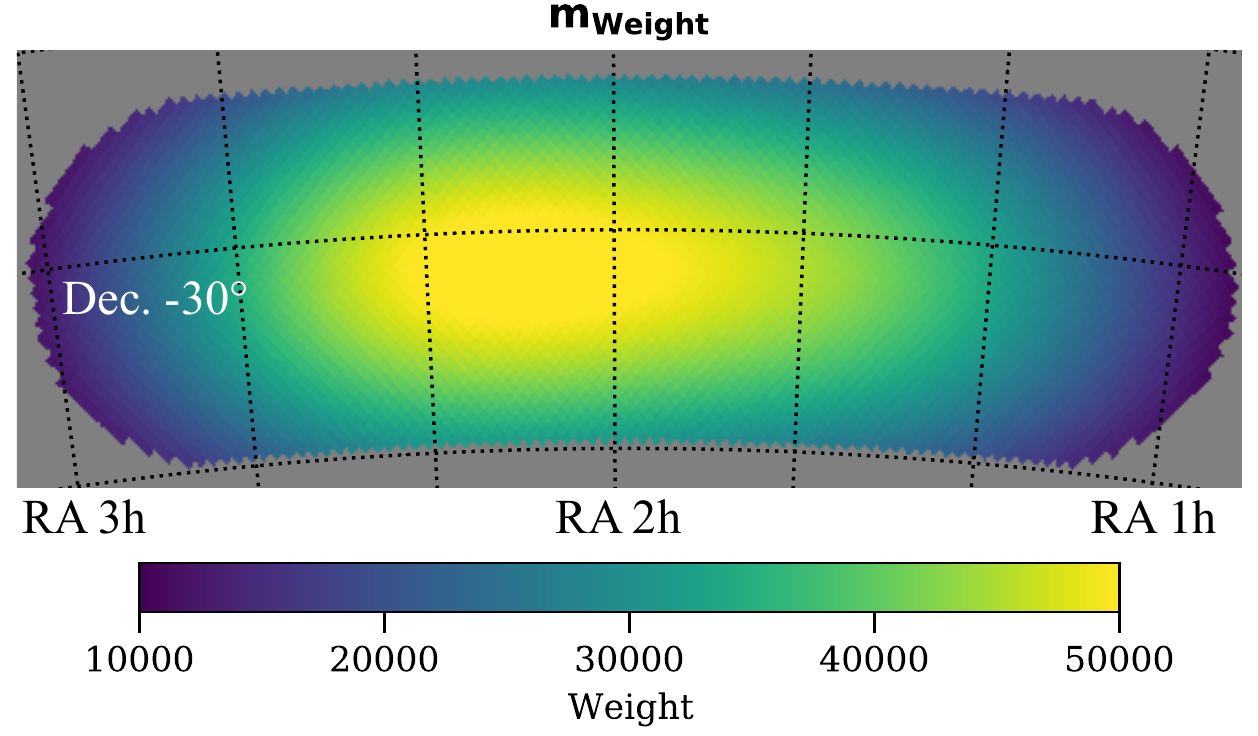}
    \caption{An example of the weight map defined in Equation~\ref{equ:map_weight}.
    The map spans two hours in RA.
    The fact that there are more effective weight on the left side of the region is due to lower noise (higher weighting) when observing that region, which results from longer integration time and lower intrinsic sky noise.
    }
    \label{fig:weight_map}
\end{figure}

We first construct another version of $\mathbf{A}$ with only the beam term
(i.e., setting the exponential term equal to unity),
which we call $\mathbf{A}_\mathrm{B}$:
\begin{equation}
    (A_B)_{nk} = 
    B(\mathbf{\hat s}_{n,k}).
\end{equation}
We then calculate the weight map
\begin{equation}
\label{equ:map_weight}
    \mathbf{m}_\mathrm{weight} = \mathbf{A}_\mathrm{B}^\dagger \mathbf{N}^{-1} \mathbf{1},
\end{equation}
where $\mathbf{1}$ is an all-ones vector counting all visibilities.
The weight map is constructed following the exact sampling of the visibility measurement and the noise weighting.
We divide the weight map out of $\mathbf{\hat{m}}$ to turn the visibility summations to averages.
Figure~\ref{fig:weight_map} shows an example of the weight map.
 
Finally, we define the $\mathbf{D}$ matrix as
\begin{equation} \label{equ:d_definition}
    \mathbf{D}  = \mathrm{diag} (\frac{1}{m_{\mathrm{w},1}}, \frac{1}{m_{\mathrm{w},2}}, ..., \frac{1}{m_{\mathrm{w}, N_\mathrm{pixel}}}),
\end{equation}
where $m_{\mathrm{w}, i}$ is the $i$th element in $\mathbf{m}_\mathrm{weight}$.
This definition accounts for the differing contribution of the visibilities to each point on the map given the primary beam.
However, the primary beam effect, from observation, is still left in the map, meaning that sources away from the zenith are attenuated by the primary beam.
\\

With $\mathbf{A, N}$, $\mathbf{D}$ and the conditioned HERA data $\mathbf{d}$, the map estimator $\mathbf{\hat m}$ is calculated with Equation~\ref{equ:optimal_mapping}.
The PSF and the covariance matrices ($\mathbf{P}$ and $\mathbf{C}$) are also calculated with Equation~\ref{equ:p_mat} and Equation~\ref{equ:covariance}.

\section{Algorithm Validation}
\label{sec:validation}
In this section, we use simulated data to validate the direct optimal mapping algorithm.

\subsection{Map Validation}
\label{subsec:map_validation}
We generate simulated data with the HERA Phase I array configuration, through a pipeline independent of direct optimal mapping.
The simulated data are verified to be consistent with the \texttt{pyuvsim} simulation~\citep{lanman/etal:2019} to machine-precision (Kim et al., 2021 in preparation).
We did not use \texttt{pyuvsim} because our simulator is tested to be more computationally efficient.

\begin{figure}
    \centering
    \includegraphics[width=\linewidth]{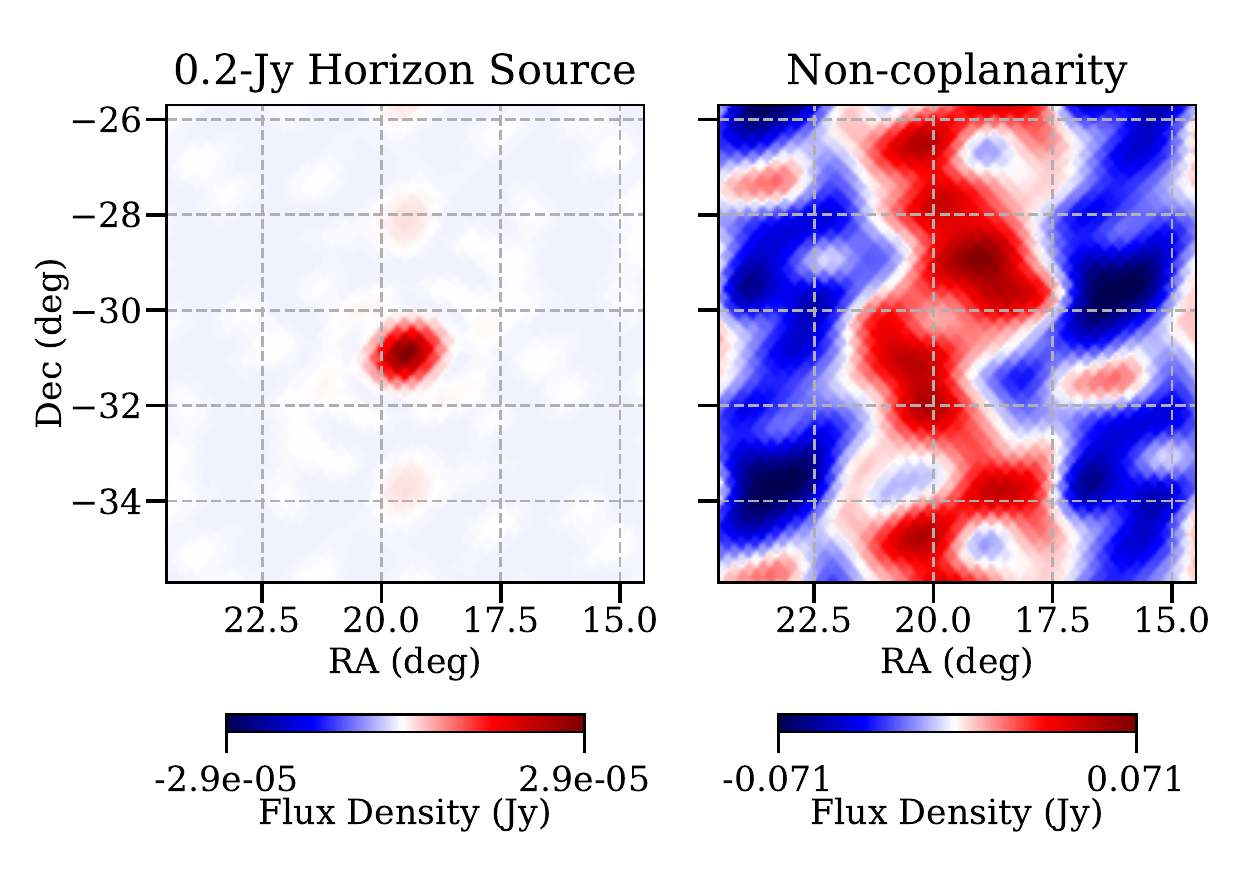}
    \caption{Validation-characterized effects: horizon source and array coplanarity. Both maps are centered on the zenith.
    \textit{Left}: A point source, with a typical flux density (0.2\,Jy at 166\,MHz), is added 85\dg{} away from the zenith.
    The source signal is picked up by the sidelobes of the primary beam and leaks into the zenith field via grating lobes at the $10^{-4}$ level.
    \textit{Right}: Non-Coplanarity residual.
    Comparing to the original map in Figure~\ref{fig:map_validation_noisy}, the difference increases to 5\% of the original map with the GSM08 sky model.
    }
    \label{fig:horizon_source_and_w-term}
\end{figure}

\begin{figure}
    \centering
    \includegraphics[width=\linewidth]{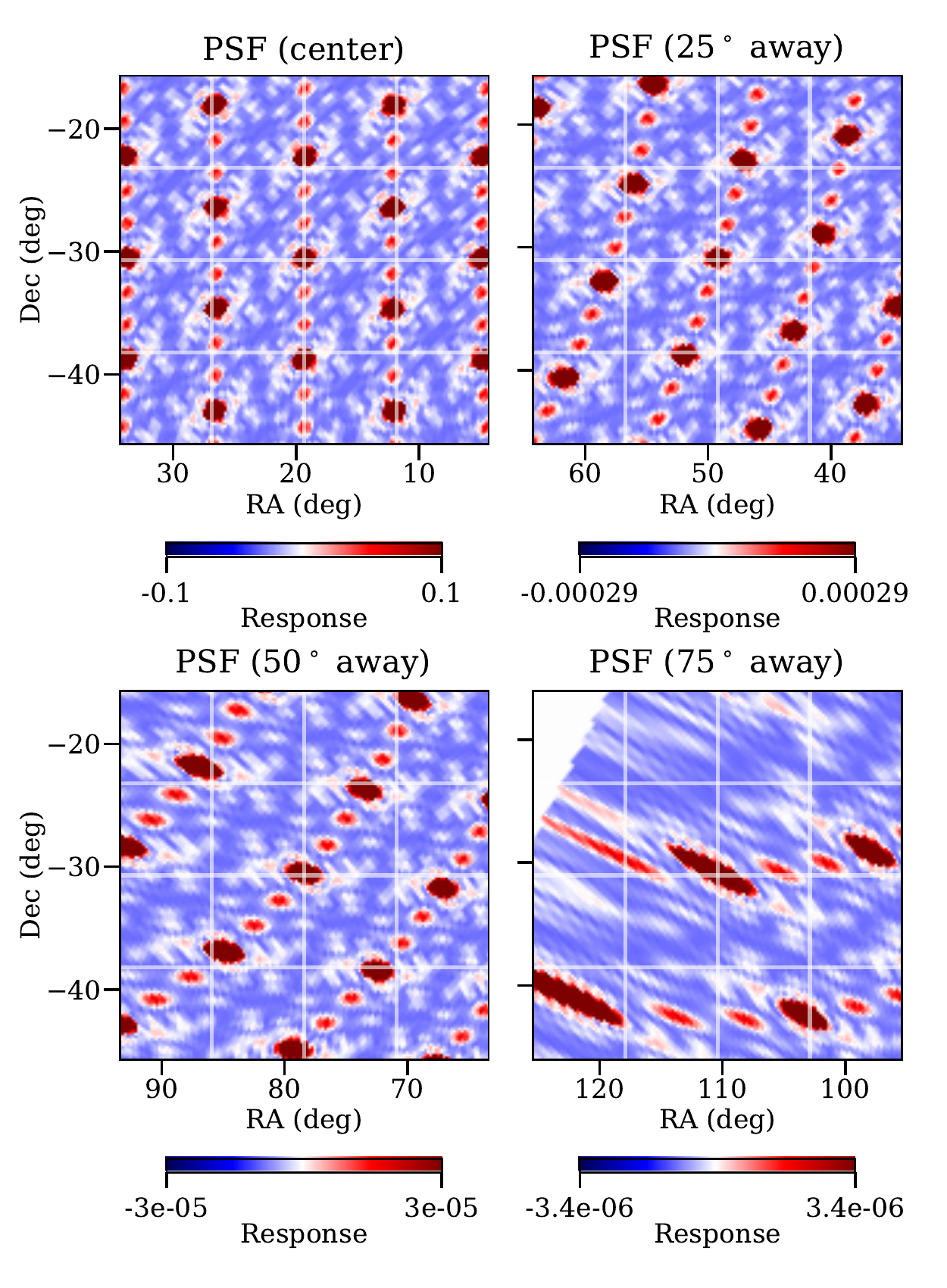}
    \caption{Point spread function (PSF) in 30\dg{}\,$\times$\,30\dg{} maps at 166\,MHz. The four maps show the field center and 25\dg{}, 50\dg{}, 75\dg{} away from the field center.
    The white area in the 75\dg{} plot is beyond the horizon.
    The full-width-at-half-maximum (FWHM) of the synthesized main beam at the field center is 50--60 arcminutes, close to the diffraction limit defined by the longest baseline. 
    Grating lobes are seen in hexagonal patterns, which are distorted as we move away from the field center, especially in the 75\dg{}-away map.
    As shown, DOM provides direction-dependent PSF information across the sky.
    The grating lobes cause the difference between the original and the convolved GSM08 in Figure~\ref{fig:map_validation_noisy}.
    The peak values change because of the primary-beam attenuation at the four pixels.
    The color bars saturate at 10\% of the peak values to illustrate faint sidelobe structures.
    }
    \label{fig:psf}
\end{figure}

\begin{figure}
    \centering
    \includegraphics[width=\linewidth]{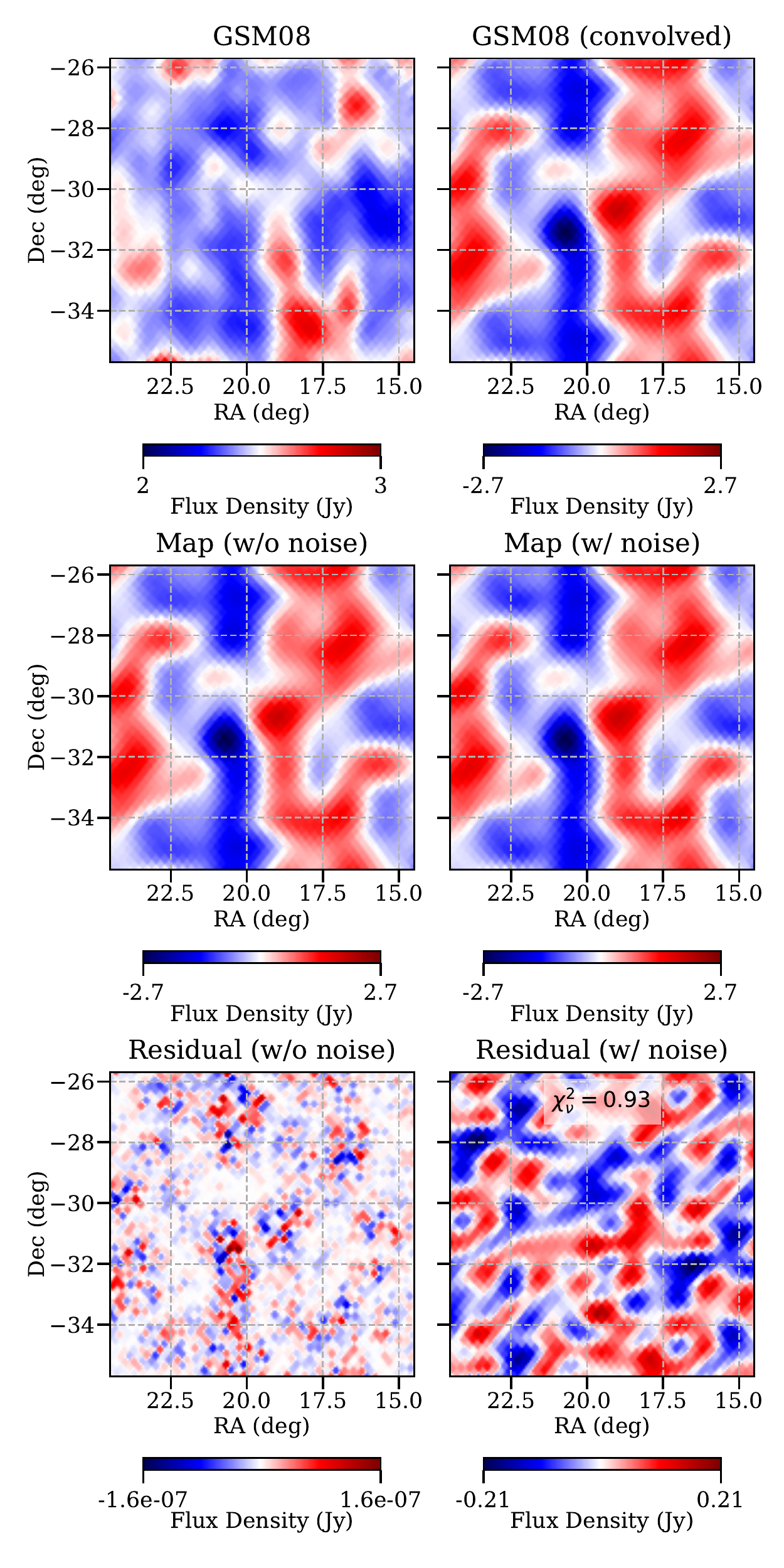}
    \caption{Map validation with simulated visibilities.
    \textit{Top}: Global Sky Model (GSM08)~\citep{doc/etal:2008} at 166\,MHz before and after convolving with the PSF.
    Because of the grating lobes in the PSF (Figure~\ref{fig:psf}), the convolved GSM08 looks different from the original GSM08.
    \textit{Middle}: the DOM map from simulated visibilities before and after adding noise.
    \textit{Bottom}: the residual maps after subtracting the convolved GSM08 from the \textit{middle} row.
    The one without noise shows $10^{-7}$ residuals. 
    The one with noise shows random patterns at $\sim$\,10\% level.
    The reduced-$\chi^2$ value (Section~\ref{subsec:covariance}) is also shown in the noise residual.
    }
    \label{fig:map_validation_noisy}
\end{figure}

We use the Global Sky Model (GSM08)~\citep{doc/etal:2008} as our model of the true sky for one twenty-second integration at 166\,MHz.
Then we use DOM to map the simulated visibilities, and compare the map to convolved GSM08.
Specifically, the convolved GSM08 is calculated by multiplying the \mb{P} matrix and the GSM08 sky model.
Without noise, the two maps are consistent at $10^{-7}$ levels (Figure~\ref{fig:map_validation_noisy}).
Getting to the $10^{-7}$ consistency, we quantitatively characterized various factors that affect the mapping results:
\begin{enumerate}
    \item Primary beams: the residual map is very sensitive to the primary beam.
    We use the CST-simulated single-antenna primary beam~\citep{fagnoni/etal:2021}, which describes the beam pattern in electric fields with 1\dg{} angular grid and 0.5\,MHz frequency resolution.
    However, our mapping requires  $<$\,0.25\dg{} angular resolution and 0.1\,MHz frequency resolution.
    We need to interpolate the simulated beam pattern in both angular and frequency space.
    We use the \texttt{UVBeam} object within the \texttt{pyuvdata} package~\citep{hazelton/etal:2017} to perform the interpolation.
    When the beam pattern is rotated by 90$^\circ$, the residual increases up to 30\% of the original map. 
    Moreover, interpolating in electric fields or in power also lead to $10^{-4}$ differences in the residual maps.
    This sensitivity to the primary beam indicates the possibility to constrain the primary beams with DOM maps.
    
    \item Horizon contribution: the simulation includes all signals above horizon. 
    We found that if we only convolve sky signals within 50\dg{} around the zenith, the residual is 30\,--\,50\% of the peak value.
    Furthermore, Figure~\ref{fig:horizon_source_and_w-term} (on the left) shows the map difference when a typical point sources is added near the horizon (85\dg{} away from the zenith).
    The source leaks into the zenith field at $10^{-4}$ level. 
    Considering the significant solid angle around the horizon, this demonstrates the necessity to correctly include signals within the entire observable hemisphere~\citep{pober/etal:2016}, perhaps even accounting for the terrain near the horizon~\citep{bassett/etal:2021}, to model the foreground.
    
    \item Coplanarity: we estimate the effect of array coplanarity by comparing the mapping results with and without assuming the antennas are on a plane.  The HERA dishes deviate randomly from a perfect plane by about 4\,cm, and
    Figure~\ref{fig:horizon_source_and_w-term} (on the right) shows the resulting 5\%-level difference from ignoring the non-coplanarity.
\end{enumerate}

DOM calculates the direction-dependent PSFs across the field.
Figure~\ref{fig:psf} shows PSFs in four pixels from the field center to near-horizon.
The synthesized beam and the grating lobe pattern become increasingly distorted as the pixel moves away from the field center, illustrating the importance of considering direction-dependence PSFs.

After mapping the noiseless simulation, we add noise to the visibilities.
Specifically for a simulated visibility $d_n$, we draw independent random noise with the amplitude of $\sigma_{n}/{\sqrt{2}}$ for the real and imaginary parts.
The noisy visibilities are then mapped to compare with the convolved sky model.
Figure~\ref{fig:map_validation_noisy} shows the results with and without adding noise to the simulations.
Without noise, the residual map is at $10^{-7}$ levels.
With noise, the residual map amplitude is $\sim$\,10\%, six orders of magnitude higher than the noiseless residual map.
Further investigation shows that the residual pattern changes for each random noise realization, confirming that it is random-noise dominated.

\subsection{Pixel Covariance Matrix}
\label{subsec:covariance}

The coherent noise pattern (Figure~\ref{fig:map_validation_noisy}, bottom right) shows the correlated noise in map space.
One feature of DOM is that it provides a robust noise covariance matrix $\mathbf{C}$ to capture the correlation.
The pixel correlation results from the instrument's incomplete $uvw$ space sampling. 
There are patterns in the sky to which the array configuration is not sensitive.

\begin{figure}
    \centering
    \includegraphics[width=\linewidth]{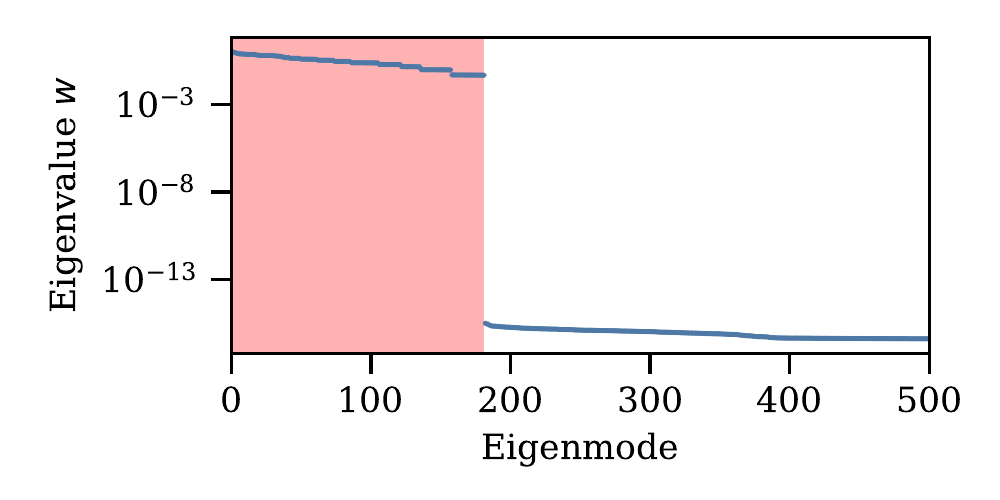}
    \caption{Eigenvalues of \mb{C} in descending order for the one-integration simulation data. 
    A clear drop is seen around the 180th eigenmode.
    Thus, we choose the first 180 eigenvalues for the figure-of-merit calculation.
    The red-shaded region shows the selected eigenmodes.
    }
    \label{fig:sim_eign}
\end{figure}

The noise covariance matrix $\mathbf{C}$ is closely related to the PSF matrix $\mathbf{P}$ by multiplying $\mathbf{D}^\mathrm{T}$ on the right (Equation~\ref{equ:covariance}).
With \mb{C}, we define a maximum likelihood figure-of-merit (FoM) to evaluate residual maps using the pixel covariance:
\begin{equation}
    \label{equ:fom}
    \mathrm{FoM} = \mathbf{\Delta m^T} \mathbf{C^{-1}}  \mathbf{\Delta m},
\end{equation}
where $\mathbf{\Delta m}$ is the map vector of one residual map.
Since $\mathbf{C}$ is not invertible, we first eigendecompose \mb{C}
\begin{equation}
\label{equ:eigendecompose}
    \mathbf{C} = \mathbf{V\, W\, V^{-1}}
\end{equation}
where $\mathbf{V}$ and $\mathbf{W}$ are the eigenvectors and eigenvalues.
Since \mb{C} is singular, \mb{W} has zero elements.
Figure~\ref{fig:sim_eign} shows the eigenvalues in descending order.
A clear drop is seen around the 180th eigenvalue, this is related to the number of nonredundant modes HERA Phase I array measures.
In addition, the fact that the simulation data only has one time integration leads to the sharp drop in eigenvalue spectrum.
Eigenvalue spectra look different when we map multiple time integrations in Section~\ref{subsec:chi2}.

Because \mb{C} is symmetric, we can choose \mb{V} to be orthonormal 
\begin{equation}
\label{equ:v_orthonormal}
    \mathbf{V^{-1} = V^T}.
\end{equation}

Now we plug Equation~\ref{equ:eigendecompose} into Equation~\ref{equ:fom}
\begin{eqnarray}
\label{equ:fom_wv}
    \mathrm{FoM} &=& \mathbf{\Delta m^T (V\,W\,V^{-1})^{-1} \Delta m} \nonumber \\
    &=& \mathbf{\Delta m^T V\,W^{-1}\,V^{-1} \Delta m} \nonumber \\
    &=& \mathbf{\Delta m^T V\,W^{-1}\,V^T \Delta m} \nonumber \\
    &=& \mathbf{(\Delta m^T V) W^{-1} (\Delta m^T V)^T },
\end{eqnarray}
where \mb{\Delta m^T V} is the map projections onto the eigenvectors.
Since \mb{W} has zero values, \mb{W^{-1}} is not computable.
Therefore, we only consider the dominating eigenvalues in Equation~\ref{equ:fom_wv}.
For this simulated data, we choose the first 180 eigenvalues, indicated in Figure~\ref{fig:sim_eign}.

We write \mb{\Delta m^T V} and \mb{W} element-wise for the first dominating 180 eigenvalues
\begin{eqnarray}
    &\mathbf{\Delta m^T V} &= (\alpha_1, \alpha_2,..., \alpha_{N=180}),\\
    &\mathbf{W} &= \mathrm{diag}(w_1, w_2,...,w_{N=180}),
\end{eqnarray}
and the FoM calculation can be more clearly expressed as
\begin{equation}
    \mathrm{FoM} = \sum_i^{N=180} \frac{\alpha_i^2}{w_i}.
\end{equation}
This FoM is a $\chi^2$-statistic, and the reduced-$\chi^2$ is $\chi^2_\nu = \chi^2/\mathrm{d.o.f}$.\footnote{The degrees of freedom (d.o.f) for this simulated data is 180, the number of eigenvalues considered.}
For a residual map from noise-dominated visibilities, we expect the FoM follows a $\chi^2$ distribution with 180 degrees of freedom.
For reduced-$\chi^2$, we expect $\chi_{180}^2\sim1$.
The simulation map in Figure~\ref{fig:map_validation_noisy} is measured $\chi_{180}^2 = 0.93$.
For different noise realizations, $\chi_{180}^2 \sim 1$, validating that the covariance matrix provides an accurate description of correlated pixel noise in map space.

Back to the interferometric setting, eigenvectors represent emission patterns in the sky.
The measured visibilities are sensitive to different patterns at different levels, which is characterized by the magnitude of the eigenvalues. 
The eigenvectors with very small eigenvalues are essentially invisible to the interferometer, which should be excluded for the FoM calculation.
Therefore, by selecting the nonzero eigenvalues, we only use the emission patterns visible to the array in calculating the FoM.

\section{Mapping HERA Data}
\label{sec:mapping_hera_data}
In this section, we map a small fraction of HERA data and evaluate sky models against the HERA measurement.
\citet{aguirre/etal:2021} report a calibration bias in the calibration of this dataset, so we correct the bias by multiplying the visibilities by a factor of 1.04~\citep{hera/etal:2021} before mapping.

\subsection{HERA Map vs Sky Models}
\label{subssec:hera_map_vs_sky_models}

We select the best fraction of the HERA Phase I data, the central region of Field 1 defined in \citet{hera/etal:2021}.
This data set contains 20 twenty-second time integrations.
We randomly select one HERA frequency channel around 166\,MHz with a bandwidth of 100\,kHz and map the corresponding visibilities.
Meanwhile, we calculate the \mb{P} matrix to cover the entire hemisphere (90\dg{} from the zenith) to convolve four sky models and compare with HERA data:

\begin{figure}
    \centering
    \includegraphics[width=\linewidth]{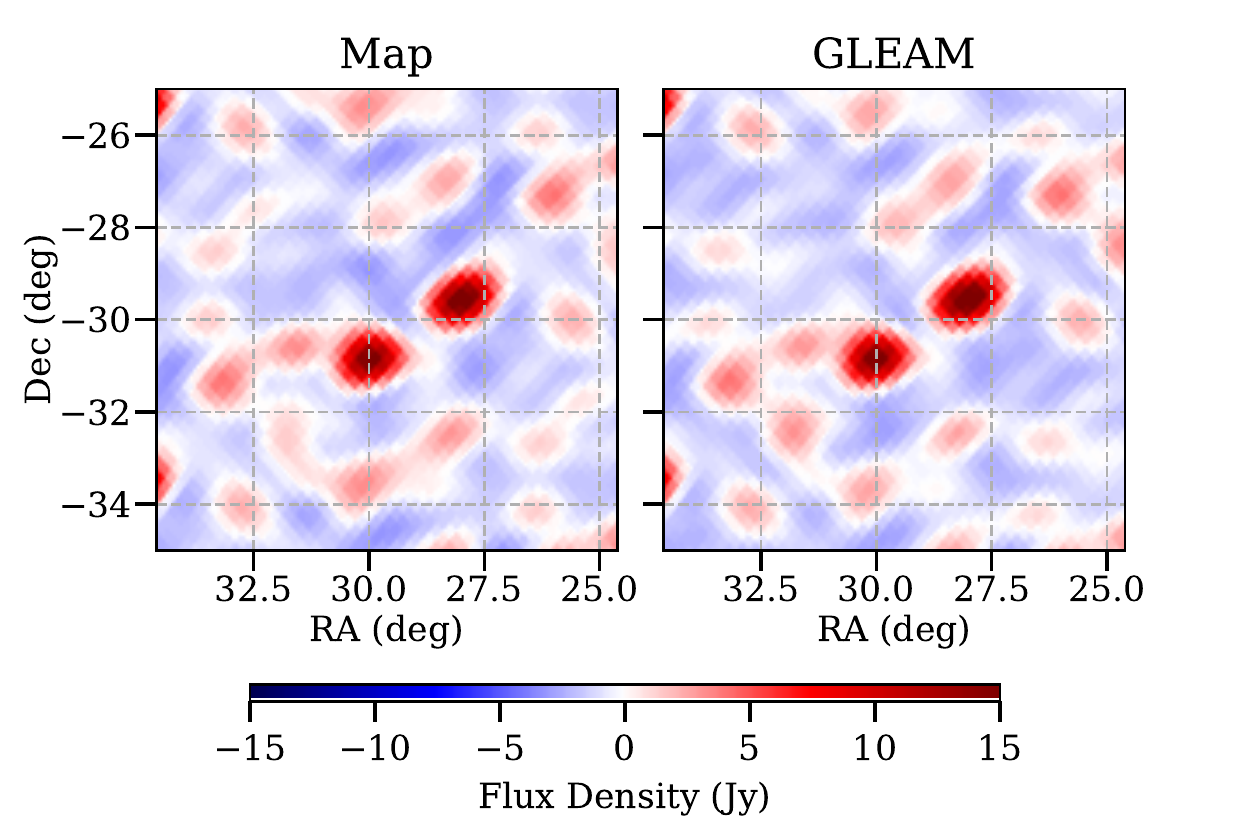}
    \caption{Maps from HERA data and convolved GLEAM catalogs~\citep{wayth/etal:2015} at 166\,MHz. 
    \textit{Left:} the HERA map. 
    \textit{Right:} GLEAM catalogs~\citep{hurley-walker/etal:2017, hurley-walker/etal:2019} convolved by the PSF.
    The GLEAM catalogs miss some diffuse emission in the map, especially between point sources. 
    }
    \label{fig:hera_data&model}
\end{figure}

\begin{figure*}
    \centering
    \includegraphics[width=\linewidth]{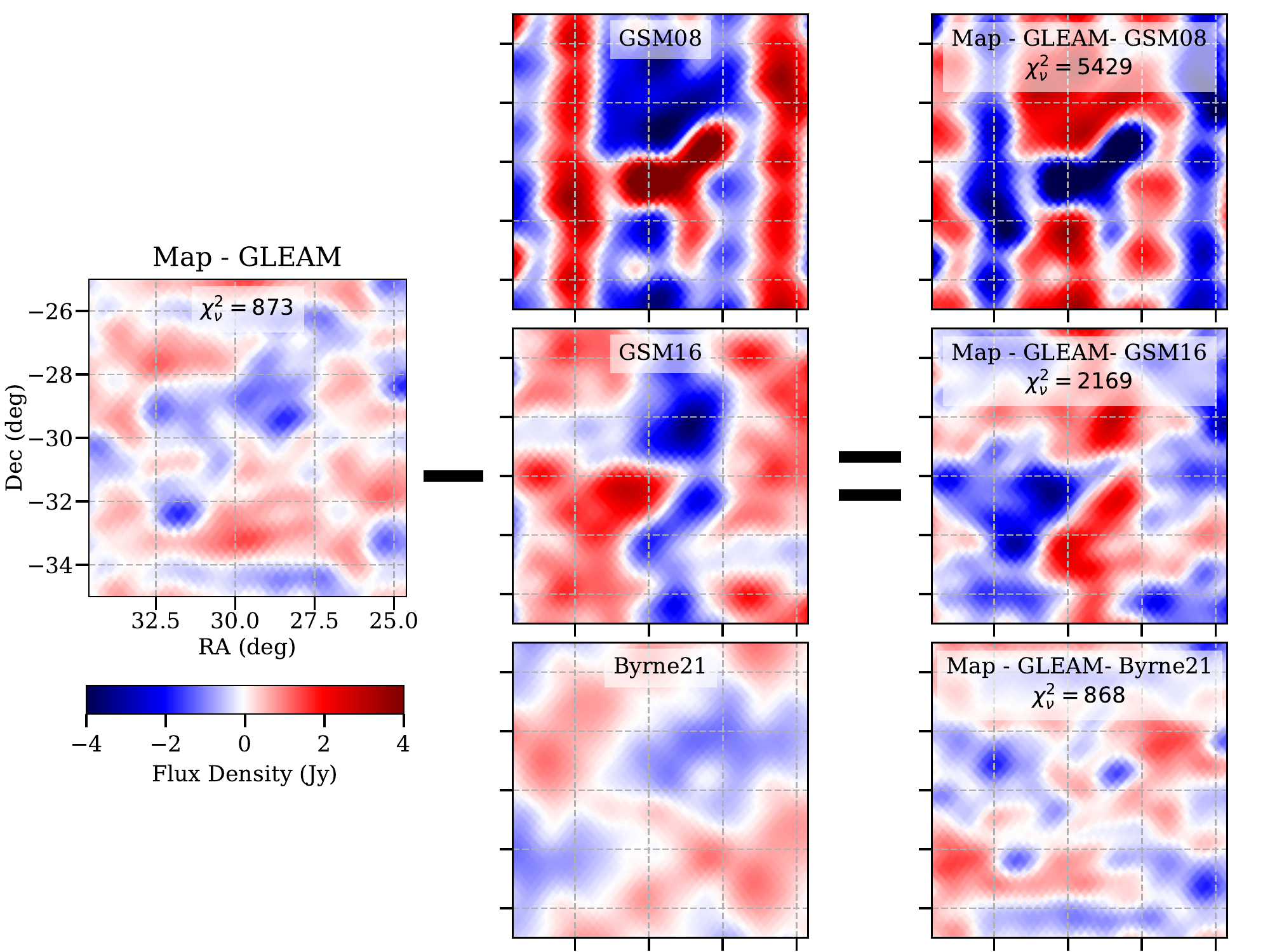}
    \caption{Residual maps at 166\,MHz.
    The figure is designed as a subtraction equation showing residuals with different components subtracted.
    All maps share the same sky patch and color bar range as annotated on the left map.
    \textit{Left column:} After subtracting the convolved GLEAM catalogs~\citep{hurley-walker/etal:2017, hurley-walker/etal:2019} from the HERA map, the residual map starts to show diffuse emission patterns.
    \textit{Center column:} From top to bottom are convolved Global Sky Model 2008 (GSM08)~\citep{doc/etal:2008}, convolved Global Sky Model 2016 (GSM16)~\citep{zheng/etal:2017}, and convolved Byrne21 map~\citep{byrne/etal:2021}.
    GSM08 contains the two brightest point sources, while neither GSM16 nor Byrne21 contains point sources.
    GSM08, less so for GSM16, also shows vertical stripes.
    \textit{Right column:} Residual maps after further subtracting the GSM08, GSM16, and Byrne21 maps.
    Both the GSM08 and the GSM16 residuals increase the amplitude of the residual map.
    The morphology of their residual maps resemble that of the sky models themselves, indicating that they do not represent the observed diffuse emission.
    The Byrne21 map, however, shows a similar diffuse pattern as observed in the GLEAM-subtracted residual at large scales.
    Reduced-$\chi^2$ values are also presented for each of the residual maps as in Table~\ref{tab:chi2}.
    }
    \label{fig:hera_residual}
\end{figure*}

\begin{enumerate}
    \item GLEAM catalogs: GLEAM~\citep{wayth/etal:2015} is a set of point source catalogs from the Murchison Widefield Array (MWA). GLEAM covers the sky south of +30\dg{} declination across 72--231\,MHz. The publicly-available data include an extragalactic catalog~\citep{hurley-walker/etal:2017} with 307,455 sources and a partial Galactic plane catalog~\citep{hurley-walker/etal:2019} with 22,037 sources.
    We use the fitted spectral information stored in the GLEAM catalog---a flux at 200\,MHz and a spectral index---for each source to evaluate its flux at 166\,MHz.
    We remove sources without fitted spectral information in GLEAM catalogs, because those sources cluster within a sky patch $>$120\dg{} away from our field.
    We do not approximate point source flux to the center of the corresponding pixel because we found that the location approximation causes significant errors.
    Instead, we create additional pixels representing the exact location of each point source, similar to what was done in \citet{dillon/etal:2015}. 
    \item Global Sky Model in 2008 (GSM08): \citet{doc/etal:2008} compiled several sky surveys and derived a model of the sky emission across 10\,MHz\,--\,94\,GHz. GSM08 includes both the diffuse emission and point sources.
    \item Global Sky Model in 2016 (GSM16): \citet{zheng/etal:2017} improved upon \citet{doc/etal:2008} by including additional or revised survey maps~\citep{remazeilles/etal:2015} and masking out the top 1\% pixels to remove point sources. 
    The frequency coverage is also extended to 10\,MHz\,--\,5\,THz from GSM08.
    \item Byrne21 Map: \citet{byrne/etal:2021} recently published a sky map at 182\,MHz from MWA observations, covering 11,000 deg$^2$.
    Assuming the Byrne21 map is dominated by Galactic synchrotron foreground, we use the reported spectral index of -2.61~\citep{mozdzen/etal:2017} to scale the original 182\,MHz map to 166\,MHz.
\end{enumerate}

The HERA map and convolved GLEAM catalogs are shown in Figure~\ref{fig:hera_data&model}.
The GLEAM catalogs match the HERA map in point-like morphology and amplitude, while missing some faint diffuse emission.
A similar comparison between GLEAM and HERA was performed in \citet{carilli/etal:2020} with the Common Astronomical Software
Applications (CASA) CLEAN imaging.

We subtract the GLEAM catalogs out of the HERA map (left plot of Figure~\ref{fig:hera_residual}) to further study the diffuse structures. 
With the bright point sources subtracted, the diffuse emission starts to emerge. 
We compare sky models with the measured diffuse emission.
Figure~\ref{fig:hera_residual} shows this comparison by further subtracting sky models out of the GLEAM-subtracted residual.
The middle column of Figure~\ref{fig:hera_residual} shows three convolved sky models.
The GSM08 model shows signs of the two bright sources and vertical stripes.
These stripes originate from the Haslam survey~\citep{haslam/etal:1982}, which is used to construct GSM08~\citep{remazeilles/etal:2015}.
GSM16 does not show signs of bight point sources nor obvious signs of vertical stripes after their de-sourcing and de-striping processes~\citep{zheng/etal:2017}.
But neither GSM08 nor GSM16 shows a diffuse emission pattern that resembles the measurement.
However, the Byrne21 map shows similar emission patterns to the measurement, especially at large scales.

After further subtracting the three sky models, the final residual maps are shown on the right column of Figure~\ref{fig:hera_residual}.
Both GSM08 and GSM16 residual maps show morphology close to GSM08 and GSM16 themselves, with negative amplitudes. This means that neither GSM08 nor GSM16 map reduces the measured diffuse pattern, instead they impose their intrinsic patterns in the final residual maps.
However, the Byrne21 residual does show reduced large-scale diffuse pattern, with the point sources more prominent in the final residual map.

\subsection{Reduced-$\chi^2$ of Residual Maps}
\label{subsec:chi2}

\begin{figure}
    \centering
    \includegraphics[width=\linewidth]{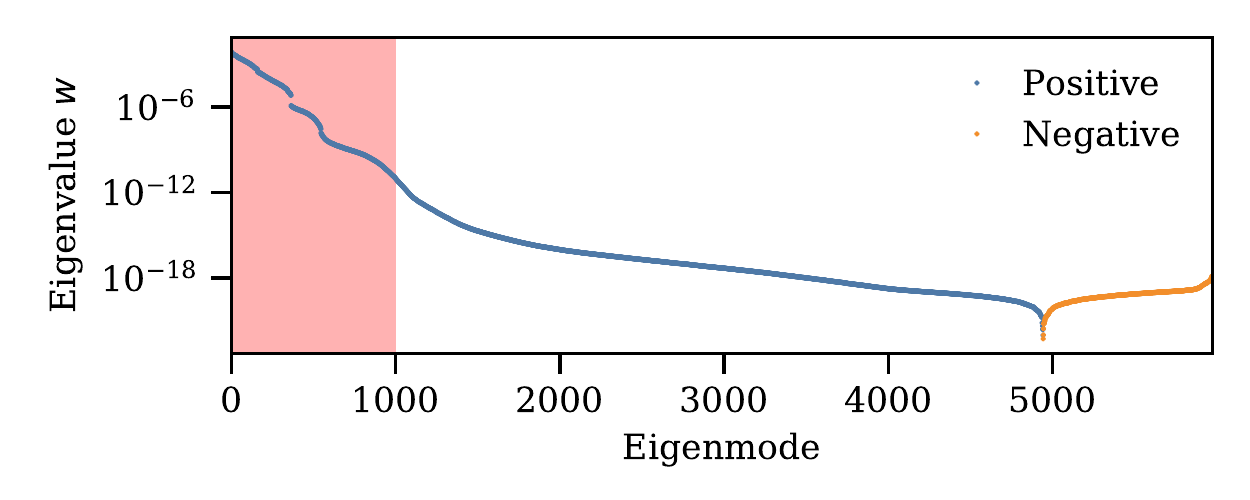}
    \caption{Full eigenvalue spectrum of \mb{C} from mapping the HERA data.
    The eigenvalues are displaced in descending order, with blue for the positive and orange for the negative.
    The negative values have the absolute values at the machine-error levels, and are artificially ordered to the right side of the plot instead of scattering around zero with the positive eigenvalues.
    Different from Figure~\ref{fig:sim_eign}, there is not a clear-cut in this eigenvalue spectrum.
    This is because we combined 20 time integrations in this mapping.
    The sky rotation adds information beyond what the same array measures at one time, smoothing the drop we saw in Figure~\ref{fig:sim_eign}.
    We choose to include the first 1000 eigenvectors (red shaded region) for the reduced-$\chi^2$ calculation.
    }
    \label{fig:data_eign}
\end{figure}

\begin{deluxetable}{lr}
\tablenum{1}
\tablecaption{Reduced-$\chi^2$ of Residual Maps\label{tab:chi2}}
\tablehead{
\colhead{Residual Map} & 
\colhead{$\chi^2_\nu$} 
}
\decimalcolnumbers
\startdata
Noise Simulation & 0.93 \\
Map - GLEAM & 873 \\
Map - GLEAM - GSM08 & 5,429 \\
Map - GLEAM - GSM16 & 2,169 \\
Map - GLEAM - Byrne21 & 868 \\
\enddata
\tablecomments{Reduced-$\chi^2$ statistics are calculated for the map from noise simulation and different residual maps.
Column (2) shows $\chi^2_\nu$ as introduced in Section~\ref{subsec:covariance}.
}
\vspace{-1cm}
\end{deluxetable}
To calculate $\chi^2_\nu$ of the maps, we first examine the eigenvalues of the covariance matrix \mb{C}.
Because we are mapping 20 time integrations, the sky rotation increases the array sampling compared to one time integration in Section~\ref{subsec:covariance}.
The additional information leads to the smoothing of the eigenvalue spectrum in Figure~\ref{fig:data_eign}.
Without a clear drop, it is not obvious to determine a cut for dominating eigenvalues.
We choose to use the first 1000 eigenvalues for this covariance matrix.
We are exploring more systematic and robust ways to determine the number of dominating eigenvalues for future work.

Similar to Section~\ref{subsec:covariance}, we measured $\chi^2_\nu$ of the residual maps in Figure~\ref{fig:hera_residual}, but with the highest 1000 eigenvalues. 
The result shows $\chi^2_{1000}$ at three levels: the \textit{Map-GLEAM} map and the \textit{Map-GLEAM-Byrne21} map show similar values at the lowest, around $0.9 \times 10^3$; the \textit{Map-GLEAM-GSM16} map shows a value in the middle, around $2 \times 10^3$; the \textit{Map-GLEAM-GSM08} map shows the highest value, around $5 \times 10^3$.

Looking at the $\chi^2_{1000}$ numbers, we can see that only
subtracting GLEAM gives a relatively low value, while further subtracting GSM08 and GSM16 increases $\chi^2_{1000}$ back up.
However, further subtracting Byrne21 does not significantly change the $\chi^2_{1000}$ value.
This numerical indication is consistent with the conclusion we visually drew in Figure~\ref{fig:hera_residual}.
The overall $\chi^2_{1000} \gg 1$ values quantify the apparent differences between HERA measurement and the sky models.
Contribution factors to the difference include errors in HERA's primary beam modeling~\citep{fagnoni/etal:2021}, in instrument calibration~\citep{dillon/etal:2020, kern/etal:2019, kern/etal:2020, kern/etal:2020b}, and in sky models~\citep{wayth/etal:2015,doc/etal:2008, zheng/etal:2017, byrne/etal:2021}.

\section{Computation Cost}
\label{sec:computation}

\begin{figure}
    \centering
    \includegraphics[width=\linewidth]{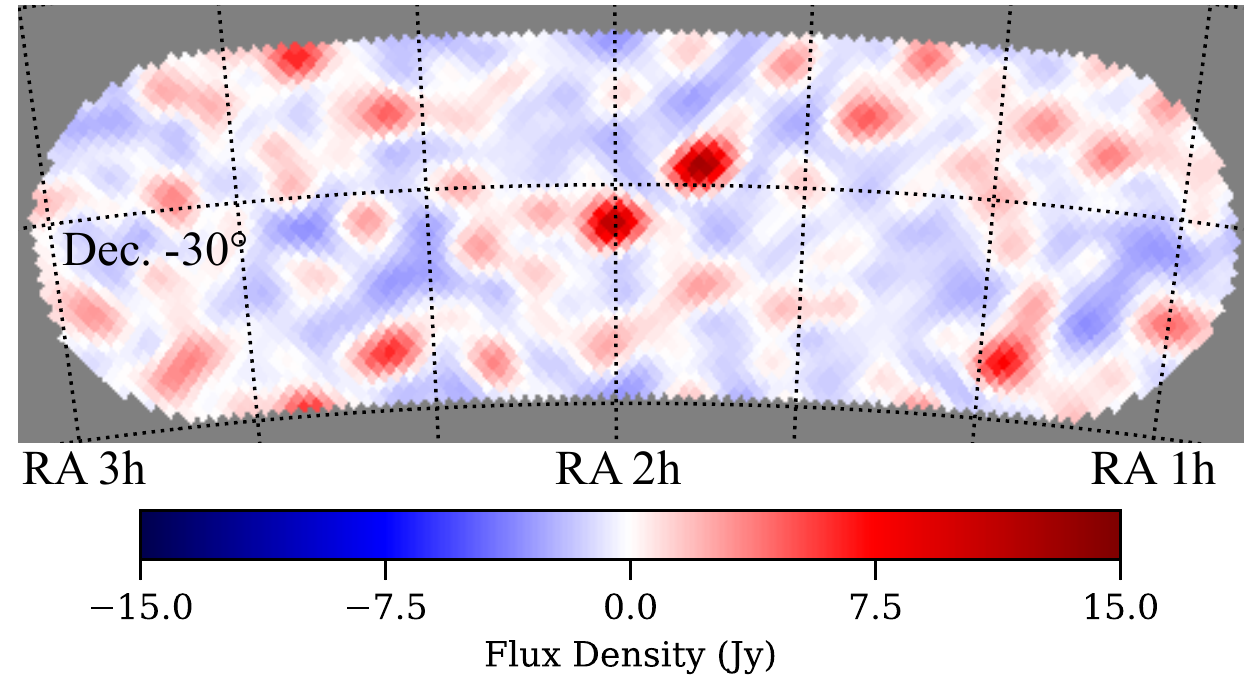}
    \caption{A wide-field map at 166\,MHz.
    Observations across RA 1h--3h are composed for this map.
    This map covers 5\dg{} around the HERA zenith path.
    Previous examples cover the central part of this field.
    }
    \label{fig:stitched_map}
\end{figure}

Previous analysis focuses on a map from 20 twenty-second integrations.
Figure~\ref{fig:stitched_map} shows a wide-field map at 166\,MHz from 300 twenty-second integrations, including the entire Field\,1~\citep{hera/etal:2021}.
This map covers 215\,deg$^2$, 5\dg{} around the RA 1\,h\,--\,3\,h sky path at -30.7\dg{} declination.
The covariance matrix for pixels in this map is also available.
Wide-field maps will be our final product. 
With the wide-field maps, \citet{liu/zhang/parsons:2016} showed that power spectra, especially in large angular scales, can be measured with spherical Fourier-Bessel formalism.
However, mapping  wide-field maps is computationally expensive.
With this mapping configuration, we investigate whether HERA Phase I data, and eventually the full HERA array data, are computable with DOM.

Here we discuss mapping within 5\dg{} around the zenith, which seems to contradict Section~\ref{subsec:map_validation} where we emphasized the contribution from the entire hemisphere, especially near the horizon.
In fact, when we map the visibilities, including a larger fraction of the sky does not change the pixel flux already mapped in a small field.
Considering the attenuation of the primary beam, the visibilities contain information mainly within a small field around the zenith.
However, if we attempt to compare the map with a sky model, we need to convolve the sky model over the entire hemisphere, as we did in Section~\ref{subssec:hera_map_vs_sky_models}.
This will be relevant for foreground removal in future cosmological analyses.

Even for a small field around the zenith, the DOM algorithm is computationally expensive both in terms of CPU time and RAM consumption.
For one frequency bin, the $\mathbf{A}$ matrix has the dimension of $N_\mathrm{visibility} \times N_\mathrm{pixel}$.
$N_\mathrm{visibility}$ is of the order of $10^{4}-10^5$, considering baselines at multiple integrations; $N_\mathrm{pixel}$ is of the order of $10^5$ or more, depending on pixel size and sky coverage.
Storing the entire matrices requires $10^1-10^2$ gigabytes of memory. 

However, since the \mb{N} matrix and the \mb{D} matrix we adopt are both diagonal, we can analyze the visibilities piecewise and add their contribution to the final map and covariance matrix.
This approach will reduce the RAM requirement.

We first write the mapping equation \mb{\hat{m} = D A^\dagger N^{-1} d} element-wise, focusing on mapping the $k$th pixel
\begin{equation}
\label{equ:mapping_sum}
    \hat{m}_k = D_k \sum_{t=1}^{N_{\mathrm{int}}} \sum_{n=1}^{N_{\mathrm{bl}}}  d_{n, t} \frac{B(\mathbf{\hat s}_{t, k})}{\sigma_{n,t}^2}  \exp \left( i \frac{2\pi\nu}{c} \mathbf{b}_n \cdot \mathbf{\hat{s}}_{t, k} \right),
\end{equation}
where $n$ runs over all the baselines and $t$ covers all the time integrations, $d_{n, t}$ is the data, and $\sigma_{n, t}$ is the noise.
The above equation shows that mapping essentially sums all the visibilities together with different coefficients that are independent of other visibilities (i.e. there are no cross terms).
The inner summation sums all the visibilities within one integration, and the outer summation sums all the time integrations.
The reason that the simple format can be obtained is that both \mb{N} and \mb{D} are diagonal.

Now we take a close look at diagonal elements in \mb{D}
\begin{equation}
\label{equ:d_k}
    D_k = \left( \sum_{t=1}^{N_{\mathrm{int}}} \sum_{n=1}^{N_{\mathrm{bl}}} \frac{B(\mathbf{\hat s}_{t, k})}{\sigma_{n, t}^2} \right)^{-1},
\end{equation}
which also sums over baselines and time integrations.
The subject to be summed is the inverse-variance weighted primary beam (Section~\ref{subsec:map_normalization}).

We plug Equation~\ref{equ:d_k} into the Equation~\ref{equ:mapping_sum}
\begin{equation}
    \hat{m}_k = \frac{\sum_{t=1}^{N_{\mathrm{int}}} \sum_{n=1}^{N_{\mathrm{bl}}}  d_{n, t} \frac{B(\mathbf{\hat s}_{t, k})}{\sigma_{n,t}^2}  \exp \left( i \frac{2\pi\nu}{c} \mathbf{b}_n \cdot \mathbf{\hat{s}}_{t, k} \right)}
    {\sum_{t=1}^{N_{\mathrm{int}}} \sum_{n=1}^{N_{\mathrm{bl}}} \frac{B(\mathbf{\hat s}_{t, k})}{\sigma_{n, t}^2}}.
\end{equation}
The numerator and denominator are completely separate, we can calculate them individually before the division. 
Furthermore, since there are no cross terms, the summations can also be performed visibility-by-visibility, avoiding large matrix multiplication.

Similarly, we write down the calculation of \mb{C} element-wise
\begin{widetext}
\begin{eqnarray}
\label{equ:c_wide_field}
    C_{k_1 k_2} &=& \frac{\sum_{t=1}^{N_{\mathrm{int}}} \sum_{n=1}^{N_{\mathrm{bl}}}  \frac{ B(\mathbf{\hat s}_{t, k_1}) B(\mathbf{\hat s}_{t, k_2})}{\sigma_{n,t}^2}  \exp \left( i \frac{2\pi\nu}{c} \mathbf{b}_n \cdot [\mathbf{s}_{t, k_1} - \mathbf{s}_{t, k_2}] \right)}
    {\left( \sum_{t=1}^{N_{\mathrm{int}}} \sum_{n=1}^{N_{\mathrm{bl}}} \frac{B(\mathbf{\hat s}_{t, k_1})}{\sigma_{n, t}^2} \right)
    \cdot
    \left( \sum_{t=1}^{N_{\mathrm{int}}} \sum_{n=1}^{N_{\mathrm{bl}}} \frac{B(\mathbf{\hat s}_{t, k_2})}{\sigma_{n, t}^2}\right) } \\
    &=& \frac{\sum_{t=1}^{N_{\mathrm{int}}} \sum_{n=1}^{N_{\mathrm{bl}}}  \frac{ B(\mathbf{\hat s}_{t, k_1}) B(\mathbf{\hat s}_{t, k_2})}{\sigma_{n,t}^2}  \exp \left( i \frac{2\pi\nu}{c} \mathbf{b}_n \cdot [\mathbf{s}_{t, k_1} - \mathbf{s}_{t, k_2}] \right)}
    {D_{k_1} \cdot D_{k_2}}.
\end{eqnarray}
\end{widetext}
Although the equation is complicated, the logic is clear: the numerator can be added visibility-by-visibility because there are no cross terms, and the denominator is the product of two diagonal elements of \mb{D}.

In practice, we do not really analyze the visibilities one-by-one, we analyze visibilities in manageable batches, say per time integration.
This operation loosens the requirement for \mb{N}: as long as there are no \textit{inter-batch} visibility noise correlation, the calculation can still be conducted piecewise, although the calculation may not be written down as simply as illustrated above.

To summarize, provided we keep track of the normalization, the final results from this piece-wise calculation are the same as if we analyzed all the visibilities simultaneously.
This piece-wise calculation frees us from storing one large $\mathbf{A}$ matrix in RAM; we are now only limited by storing the noise covariance matrix \mb{C}.
The summations in the piecewise calculation also shows that the computation scales linearly with $N_\mathrm{baseline}$ and $N_\mathrm{integration}$.
Since the expressions are for one pixel or one matrix element, the computation scales with $N_\mathrm{pixel}$ for the maps while scales with $N_\mathrm{pixel}^2$ for the noise covariance matrix.
At last, the above discussion is about mapping one frequency channel, the computation also scales linearly with number of frequency channels.

\begin{deluxetable}{rrrrrr}
\tablenum{2}
\tablecaption{DOM Computation\label{tab:computation}}
\tablehead{
\multicolumn{6}{c}{HERA Phase I (Nside=256)} \\
\colhead{Baseline} & \colhead{Int.} & \colhead{Pixel} & \colhead{Freq.} & \colhead{CPU time} & \colhead{RAM}
}
\startdata
595 & 1 & 5411 & 1 & 38\,sec & 0.7\,GB\\
595 & 300 & 5411 & 1 & 2.8\,hrs & 0.7\,GB\\
595 & 300 & 5411 & 300 & 840\,hrs & 0.7\,GB\\
\hline
\hline
\multicolumn{6}{c}{HERA 320-Antenna (Nside=512)}\\
\colhead{Baseline} & \colhead{Int.} & \colhead{Pixel} & \colhead{Freq.} & \colhead{CPU time} & \colhead{RAM}\\
\hline
1501 & 1 & 21644 & 1 & 380\,sec & 10\,GB\\
1501 & 300 & 21644 & 1 & 28\,hrs & 10\,GB\\
1501 & 300 & 21644 & 300 & 8400\,hrs & 10\,GB\\
\enddata
\tablecomments{DOM computation with different mapping parameter, including the numbers of baselines, time integrations, pixels, and frequency channels.
Baseline numbers shown are non-redundant averaged number for HERA Phase I and redundant averaged number for HERA 320-antenna.
The calulation includes mapping and calculating the pixel covariance matrices.
The quoted time is for single-core CPU time for Intel Xeon CPU E5-2643 v2 @ 3.50\,GHz.}
\vspace{-1cm}
\end{deluxetable}

Table~\ref{tab:computation} shows the computation time for mapping the HERA Phase I data and the upcoming HERA data with 320 antennas.
In summary, it takes 2.8 single-core CPU hours and 0.7\,GB of RAM to generate the map in Figure~\ref{fig:stitched_map}, along with the covariance matrix.
Computing maps across 300 frequency channels takes 840 single-core CPU hours, which can be done in one day with a 40-core server.
Projecting to the 320-antenna HERA data, we will map 2.5 times more baselines and four times more pixels (from increasing map resolution).
It will take 28 single-core CPU hours and 10\,GB to calculate one map along with the covariance matrix.
Calculating 300 frequency channels requires 8,400 single-core CPU hours, which can be done in $<9$\,days with a 40-core server.

\section{Conclusion}
\label{sec:conslusion}
We have introduced the direct optimal mapping algorithm for interferometric data and applied the algorithm to HERA Phase~I data.
The algorithm is designed to recover faint diffuse emission on a wide-field with well-defined statistics.
It relaxes the requirements of small fields and coplanar antennas, which will be useful for future interferometer arrays, including space missions.

The algorithm is validated with simulated data for one twenty-second integration, showing
that it achieves $10^{-7}$ precision with noiseless data.
After adding noise, we show that, with the noise covariance matrix, the map noise follows $\chi^2_\nu \sim 1$ distribution.
The $\chi_\nu^2$ calculation accounts for the pixel correlation, which is essential for an interferometer array with incomplete $uvw$ coverage like HERA.

After correcting the calibration bias~\citep{hera/etal:2021}, we map HERA Phase I data and compare the map with the GLEAM point source
catalog and 
three different global sky models. 
The GLEAM catalogs~\citep{hurley-walker/etal:2017, hurley-walker/etal:2019} match the point sources in the map.
After subtracting the GLEAM point sources, the residual diffuse emission is best represented by the Byrne21 map~\citep{byrne/etal:2021}.
Neither GSM08~\citep{doc/etal:2008} nor GSM16~\citep{zheng/etal:2017} provides a sky emission distribution consistent with the HERA measurement.

Finally, we presented an example of wide-field mapping along with calculating the pixel covariance matrix.
We examined the computation cost and found that, with diagonal visibility noise matrix \mb{N} and normalization matrix \mb{D}, the mapping and covariance calculation can be conducted in batches of visibilities.
This reduces the memory requirement for manipulating all visibility data at once.
In addition to memory, the computation cost for mapping the HERA Phase I data and 320-antenna HERA data is one day and one week with a modern server.

With the direct optimal mapping algorithm, we are able to map visibilities into image cubes along with covariance matrices and point spread functions.
We aim to cover followup analyses in future publications.

\section*{Acknowledgements}

We thank Bang Nhan for useful discussions.
This material is based upon work supported by the National Science Foundation under Grant Nos. 1636646 and 1836019 and institutional support from the HERA collaboration partners.
This research is funded in part by the Gordon and Betty Moore Foundation through grant GBMF5215 to the Massachusetts Institute of Technology.
HERA is hosted by the South African Radio Astronomy Observatory, which is a facility of the National Research Foundation, an agency of the Department of Science and Innovation.
Parts of this research were supported by the Australian Research Council Centre of Excellence for All Sky Astrophysics in 3 Dimensions (ASTRO 3D), through project number CE170100013.
K.-F. Chen acknowledges the funding support from the Taiwan Think Global Education Trust Fellowship.
G.~Bernardi acknowledges funding from the INAF PRIN-SKA 2017 project 1.05.01.88.04 (FORECaST), support from the Ministero degli Affari Esteri della Cooperazione Internazionale - Direzione Generale per la Promozione del Sistema Paese Progetto di Grande Rilevanza ZA18GR02 and the National Research Foundation of South Africa (Grant Number 113121) as part of the ISARP RADIOSKY2020 Joint Research Scheme, from the Royal Society and the Newton Fund under grant NA150184 and from the National Research Foundation of South Africa (grant No. 103424).
P.~Bull acknowledges funding for part of this research from the European Research Council (ERC) under the European Union's Horizon 2020 research and innovation programme (Grant agreement No. 948764), and from STFC Grant ST/T000341/1.
E. ~de Lera Acedo acknowledges the funding support of the UKRI Science and Technology Facilities Council SKA grant.
J.S.~Dillon gratefully acknowledges the support of the NSF AAPF award \#1701536.
N.~Kern acknowledges support from the MIT Pappalardo fellowship.
A.~Liu acknowledges support from the New Frontiers in Research Fund Exploration grant program, the Canadian Institute for Advanced Research (CIFAR) Azrieli Global Scholars program, a Natural Sciences and Engineering Research Council of Canada (NSERC) Discovery Grant and a Discovery Launch Supplement, the Sloan Research Fellowship, and the William Dawson Scholarship at McGill.
We gratefully acknowledge the anonymous referees whose suggestions significantly improved the clarity of this work.

\bibliography{main}{}

\begin{thebibliography}{}
\expandafter\ifx\csname natexlab\endcsname\relax\def\natexlab#1{#1}\fi
\providecommand{\url}[1]{\href{#1}{#1}}
\providecommand{\dodoi}[1]{doi:~\href{http://doi.org/#1}{\nolinkurl{#1}}}
\providecommand{\doeprint}[1]{\href{http://ascl.net/#1}{\nolinkurl{http://ascl.net/#1}}}
\providecommand{\doarXiv}[1]{\href{https://arxiv.org/abs/#1}{\nolinkurl{https://arxiv.org/abs/#1}}}

\bibitem[{{Aguirre} {et~al.}(2021){Aguirre}, {Murray}, {Pascua}, {Martinot},
  {Burba}, {Dillon}, {Jacobs}, {Kern}, {Kittiwisit}, {Kolopanis}, {Lanman},
  {Liu}, {Whitler}, {Abdurashidova}, {Alexander}, {Ali}, {Balfour},
  {Beardsley}, {Bernardi}, {Billings}, {Bowman}, {Bradley}, {Bull}, {Carey},
  {Carilli}, {Cheng}, {DeBoer}, {Dexter}, {de Lera Acedo}, {Ely}, {Ewall-Wice},
  {Fagnoni}, {Fritz}, {Furlanetto}, {Gale-Sides}, {Glendenning}, {Gorthi},
  {Greig}, {Grobbelaar}, {Halday}, {Hazelton}, {Hewitt}, {Hickish}, {Julius},
  {Kerrigan}, {Kohn}, {La Plante}, {Lekalake}, {Lewis}, {MacMahon}, {Malan},
  {Malgas}, {Maree}, {Matsetela}, {Mesinger}, {Molewa}, {Morales}, {Mosiane},
  {Neben}, {Nikolic}, {Parsons}, {Patra}, {Pieterse}, {Pober}, {Razavi-Ghods},
  {Ringuette}, {Robnett}, {Rosie}, {Santos}, {Sims}, {Singh}, {Smith}, {Syce},
  {Thyagarajan}, {Williams}, \& {Zheng}}]{aguirre/etal:2021}
{Aguirre}, J.~E., {Murray}, S.~G., {Pascua}, R., {et~al.} 2021, arXiv e-prints,
  arXiv:2104.09547.
\newblock \doarXiv{2104.09547}

\bibitem[{{Bandura} {et~al.}(2014){Bandura}, {Addison}, {Amiri}, {Bond},
  {Campbell-Wilson}, {Connor}, {Cliche}, {Davis}, {Deng}, {Denman}, {Dobbs},
  {Fandino}, {Gibbs}, {Gilbert}, {Halpern}, {Hanna}, {Hincks}, {Hinshaw},
  {H{\"o}fer}, {Klages}, {Landecker}, {Masui}, {Mena Parra}, {Newburgh}, {Pen},
  {Peterson}, {Recnik}, {Shaw}, {Sigurdson}, {Sitwell}, {Smecher}, {Smegal},
  {Vanderlinde}, \& {Wiebe}}]{bandura/etal:2014}
{Bandura}, K., {Addison}, G.~E., {Amiri}, M., {et~al.} 2014, in Society of
  Photo-Optical Instrumentation Engineers (SPIE) Conference Series, Vol. 9145,
  Ground-based and Airborne Telescopes V, ed. L.~M. {Stepp}, R.~{Gilmozzi}, \&
  H.~J. {Hall}, 914522, \dodoi{10.1117/12.2054950}

\bibitem[{{Barry} {et~al.}(2019){Barry}, {Beardsley}, {Byrne}, {Hazelton},
  {Morales}, {Pober}, \& {Sullivan}}]{barry/etal:2019}
{Barry}, N., {Beardsley}, A.~P., {Byrne}, R., {et~al.} 2019, \pasa, 36, e026,
  \dodoi{10.1017/pasa.2019.21}

\bibitem[{{Barry} \& {Chokshi}(2022)}]{barry/chokshi:2022}
{Barry}, N., \& {Chokshi}, A. 2022, arXiv e-prints, arXiv:2203.01130.
\newblock \doarXiv{2203.01130}

\bibitem[{{Bassett} {et~al.}(2021){Bassett}, {Rapetti}, {Tauscher}, {Nhan},
  {Bordenave}, {Hibbard}, \& {Burns}}]{bassett/etal:2021}
{Bassett}, N., {Rapetti}, D., {Tauscher}, K., {et~al.} 2021, arXiv e-prints,
  arXiv:2106.02153.
\newblock \doarXiv{2106.02153}

\bibitem[{{Beardsley} {et~al.}(2016){Beardsley}, {Hazelton}, {Sullivan},
  {Carroll}, {Barry}, {Rahimi}, {Pindor}, {Trott}, {Line}, {Jacobs}, {Morales},
  {Pober}, {Bernardi}, {Bowman}, {Busch}, {Briggs}, {Cappallo}, {Corey}, {de
  Oliveira-Costa}, {Dillon}, {Emrich}, {Ewall-Wice}, {Feng}, {Gaensler},
  {Goeke}, {Greenhill}, {Hewitt}, {Hurley-Walker}, {Johnston-Hollitt},
  {Kaplan}, {Kasper}, {Kim}, {Kratzenberg}, {Lenc}, {Loeb}, {Lonsdale},
  {Lynch}, {McKinley}, {McWhirter}, {Mitchell}, {Morgan}, {Neben},
  {Thyagarajan}, {Oberoi}, {Offringa}, {Ord}, {Paul}, {Prabu}, {Procopio},
  {Riding}, {Rogers}, {Roshi}, {Udaya Shankar}, {Sethi}, {Srivani},
  {Subrahmanyan}, {Tegmark}, {Tingay}, {Waterson}, {Wayth}, {Webster},
  {Whitney}, {Williams}, {Williams}, {Wu}, \& {Wyithe}}]{beardsley/etal:2016}
{Beardsley}, A.~P., {Hazelton}, B.~J., {Sullivan}, I.~S., {et~al.} 2016, \apj,
  833, 102, \dodoi{10.3847/1538-4357/833/1/102}

\bibitem[{{Bennett} {et~al.}(1996){Bennett}, {Banday}, {Gorski}, {Hinshaw},
  {Jackson}, {Keegstra}, {Kogut}, {Smoot}, {Wilkinson}, \&
  {Wright}}]{bennett/etal:1996}
{Bennett}, C.~L., {Banday}, A.~J., {Gorski}, K.~M., {et~al.} 1996, \apjl, 464,
  L1, \dodoi{10.1086/310075}

\bibitem[{{Bennett} {et~al.}(2013){Bennett}, {Larson}, {Weiland}, {Jarosik},
  {Hinshaw}, {Odegard}, {Smith}, {Hill}, {Gold}, {Halpern}, {Komatsu}, {Nolta},
  {Page}, {Spergel}, {Wollack}, {Dunkley}, {Kogut}, {Limon}, {Meyer}, {Tucker},
  \& {Wright}}]{bennett/etal:2013}
{Bennett}, C.~L., {Larson}, D., {Weiland}, J.~L., {et~al.} 2013, \apjs, 208,
  20, \dodoi{10.1088/0067-0049/208/2/20}

\bibitem[{{Bernardi} {et~al.}(2013){Bernardi}, {Greenhill}, {Mitchell}, {Ord},
  {Hazelton}, {Gaensler}, {de Oliveira-Costa}, {Morales}, {Udaya Shankar},
  {Subrahmanyan}, {Wayth}, {Lenc}, {Williams}, {Arcus}, {Arora}, {Barnes},
  {Bowman}, {Briggs}, {Bunton}, {Cappallo}, {Corey}, {Deshpande}, {deSouza},
  {Emrich}, {Goeke}, {Herne}, {Hewitt}, {Johnston-Hollitt}, {Kaplan}, {Kasper},
  {Kincaid}, {Koenig}, {Kratzenberg}, {Lonsdale}, {Lynch}, {McWhirter},
  {Morgan}, {Oberoi}, {Pathikulangara}, {Prabu}, {Remillard}, {Rogers},
  {Roshi}, {Salah}, {Sault}, {Srivani}, {Stevens}, {Tingay}, {Waterson},
  {Webster}, {Whitney}, {Williams}, \& {Wyithe}}]{bernardi/etal:2013}
{Bernardi}, G., {Greenhill}, L.~J., {Mitchell}, D.~A., {et~al.} 2013, \apj,
  771, 105, \dodoi{10.1088/0004-637X/771/2/105}

\bibitem[{{Byrne} {et~al.}(2021){Byrne}, {Morales}, {Hazelton}, {Sullivan},
  {Barry}, {Lynch}, {Line}, \& {Jacobs}}]{byrne/etal:2021}
{Byrne}, R., {Morales}, M.~F., {Hazelton}, B., {et~al.} 2021, arXiv e-prints,
  arXiv:2107.11487.
\newblock \doarXiv{2107.11487}

\bibitem[{{Carilli} {et~al.}(2020){Carilli}, {Thyagarajan}, {Kent}, {Nikolic},
  {Gale-Sides}, {Kern}, {Bernardi}, {Mesinger}, {Matika}, {Abdurashidova},
  {Aguirre}, {Alexander}, {Ali}, {Balfour}, {Beardsley}, {Billings}, {Bowman},
  {Bradley}, {Bull}, {Burba}, {Cheng}, {DeBoer}, {Dexter}, {Acedo}, {Dillon},
  {Ewall-Wice}, {Fagnoni}, {Fritz}, {Furlanetto}, {Gale-Sides}, {Glendenning},
  {Gorthi}, {Greig}, {Grobbelaar}, {Halday}, {Hazelton}, {Hewitt}, {Hickish},
  {Jacobs}, {Josaitis}, {Julius}, {Kerrigan}, {Kim}, {Kittiwisit}, {Kohn},
  {Kolopanis}, {Lanman}, {La Plante}, {Lekalake}, {Liu}, {MacMahon}, {Malan},
  {Malgas}, {Maree}, {Martinot}, {Matsetela}, {Molewa}, {Morales}, {Mosiane},
  {Neben}, {Parra}, {Parsons}, {Patra}, {Pieterse}, {Pober}, {Razavi-Ghods},
  {Robnett}, {Rosie}, {Sims}, {Syce}, {Williams}, \&
  {Zheng}}]{carilli/etal:2020}
{Carilli}, C.~L., {Thyagarajan}, N., {Kent}, J., {et~al.} 2020, \apjs, 247, 67,
  \dodoi{10.3847/1538-4365/ab77b1}

\bibitem[{{Clark}(1980)}]{clark:1980}
{Clark}, B.~G. 1980, \aap, 89, 377

\bibitem[{{Cornwell}(2008)}]{cornwell:2008}
{Cornwell}, T.~J. 2008, IEEE Journal of Selected Topics in Signal Processing,
  2, 793, \dodoi{10.1109/JSTSP.2008.2006388}

\bibitem[{{Cornwell} {et~al.}(2005){Cornwell}, {Golap}, \&
  {Bhatnagar}}]{cornwell/golap:2005}
{Cornwell}, T.~J., {Golap}, K., \& {Bhatnagar}, S. 2005, in Astronomical
  Society of the Pacific Conference Series, Vol. 347, Astronomical Data
  Analysis Software and Systems XIV, ed. P.~{Shopbell}, M.~{Britton}, \&
  R.~{Ebert}, 86

\bibitem[{{Cornwell} {et~al.}(2012){Cornwell}, {Voronkov}, \&
  {Humphreys}}]{cornwell/voronkov/humphreys:2012}
{Cornwell}, T.~J., {Voronkov}, M.~A., \& {Humphreys}, B. 2012, in Society of
  Photo-Optical Instrumentation Engineers (SPIE) Conference Series, Vol. 8500,
  Image Reconstruction from Incomplete Data VII, ed. P.~J. {Bones}, M.~A.
  {Fiddy}, \& R.~P. {Millane}, 85000L, \dodoi{10.1117/12.929336}

\bibitem[{{de Oliveira-Costa} {et~al.}(2008){de Oliveira-Costa}, {Tegmark},
  {Gaensler}, {Jonas}, {Landecker}, \& {Reich}}]{doc/etal:2008}
{de Oliveira-Costa}, A., {Tegmark}, M., {Gaensler}, B.~M., {et~al.} 2008,
  \mnras, 388, 247, \dodoi{10.1111/j.1365-2966.2008.13376.x}

\bibitem[{{DeBoer} {et~al.}(2017){DeBoer}, {Parsons}, {Aguirre}, {Alexander},
  {Ali}, {Beardsley}, {Bernardi}, {Bowman}, {Bradley}, {Carilli}, {Cheng}, {de
  Lera Acedo}, {Dillon}, {Ewall-Wice}, {Fadana}, {Fagnoni}, {Fritz},
  {Furlanetto}, {Glendenning}, {Greig}, {Grobbelaar}, {Hazelton}, {Hewitt},
  {Hickish}, {Jacobs}, {Julius}, {Kariseb}, {Kohn}, {Lekalake}, {Liu}, {Loots},
  {MacMahon}, {Malan}, {Malgas}, {Maree}, {Martinot}, {Mathison}, {Matsetela},
  {Mesinger}, {Morales}, {Neben}, {Patra}, {Pieterse}, {Pober}, {Razavi-Ghods},
  {Ringuette}, {Robnett}, {Rosie}, {Sell}, {Smith}, {Syce}, {Tegmark},
  {Thyagarajan}, {Williams}, \& {Zheng}}]{deboer/etal:2017}
{DeBoer}, D.~R., {Parsons}, A.~R., {Aguirre}, J.~E., {et~al.} 2017, \pasp, 129,
  045001, \dodoi{10.1088/1538-3873/129/974/045001}

\bibitem[{{Dillon} \& {Parsons}(2016)}]{dillon/parsons:2016}
{Dillon}, J.~S., \& {Parsons}, A.~R. 2016, \apj, 826, 181,
  \dodoi{10.3847/0004-637X/826/2/181}

\bibitem[{{Dillon} {et~al.}(2014){Dillon}, {Liu}, {Williams}, {Hewitt},
  {Tegmark}, {Morgan}, {Levine}, {Morales}, {Tingay}, {Bernardi}, {Bowman},
  {Briggs}, {Cappallo}, {Emrich}, {Mitchell}, {Oberoi}, {Prabu}, {Wayth}, \&
  {Webster}}]{dillon/etal:2014}
{Dillon}, J.~S., {Liu}, A., {Williams}, C.~L., {et~al.} 2014, \prd, 89, 023002,
  \dodoi{10.1103/PhysRevD.89.023002}

\bibitem[{{Dillon} {et~al.}(2015{\natexlab{a}}){Dillon}, {Neben}, {Hewitt},
  {Tegmark}, {Barry}, {Beardsley}, {Bowman}, {Briggs}, {Carroll}, {de
  Oliveira-Costa}, {Ewall-Wice}, {Feng}, {Greenhill}, {Hazelton}, {Hernquist},
  {Hurley-Walker}, {Jacobs}, {Kim}, {Kittiwisit}, {Lenc}, {Line}, {Loeb},
  {McKinley}, {Mitchell}, {Morales}, {Offringa}, {Paul}, {Pindor}, {Pober},
  {Procopio}, {Riding}, {Sethi}, {Shankar}, {Subrahmanyan}, {Sullivan},
  {Thyagarajan}, {Tingay}, {Trott}, {Wayth}, {Webster}, {Wyithe}, {Bernardi},
  {Cappallo}, {Deshpande}, {Johnston-Hollitt}, {Kaplan}, {Lonsdale},
  {McWhirter}, {Morgan}, {Oberoi}, {Ord}, {Prabu}, {Srivani}, {Williams}, \&
  {Williams}}]{dillon/etal:2015b}
{Dillon}, J.~S., {Neben}, A.~R., {Hewitt}, J.~N., {et~al.} 2015{\natexlab{a}},
  \prd, 91, 123011, \dodoi{10.1103/PhysRevD.91.123011}

\bibitem[{{Dillon} {et~al.}(2015{\natexlab{b}}){Dillon}, {Tegmark}, {Liu},
  {Ewall-Wice}, {Hewitt}, {Morales}, {Neben}, {Parsons}, \&
  {Zheng}}]{dillon/etal:2015}
{Dillon}, J.~S., {Tegmark}, M., {Liu}, A., {et~al.} 2015{\natexlab{b}}, \prd,
  91, 023002, \dodoi{10.1103/PhysRevD.91.023002}

\bibitem[{{Dillon} {et~al.}(2020){Dillon}, {Lee}, {Ali}, {Parsons}, {Orosz},
  {Nunhokee}, {La Plante}, {Beardsley}, {Kern}, {Abdurashidova}, {Aguirre},
  {Alexander}, {Balfour}, {Bernardi}, {Billings}, {Bowman}, {Bradley}, {Bull},
  {Burba}, {Carey}, {Carilli}, {Cheng}, {DeBoer}, {Dexter}, {de Lera Acedo},
  {Ely}, {Ewall-Wice}, {Fagnoni}, {Fritz}, {Furlanetto}, {Gale-Sides},
  {Glendenning}, {Gorthi}, {Greig}, {Grobbelaar}, {Halday}, {Hazelton},
  {Hewitt}, {Hickish}, {Jacobs}, {Julius}, {Kerrigan}, {Kittiwisit}, {Kohn},
  {Kolopanis}, {Lanman}, {Lekalake}, {Lewis}, {Liu}, {Ma}, {MacMahon}, {Malan},
  {Malgas}, {Maree}, {Martinot}, {Matsetela}, {Mesinger}, {Molewa}, {Morales},
  {Mosiane}, {Murray}, {Neben}, {Nikolic}, {Pascua}, {Patra}, {Pieterse},
  {Pober}, {Razavi-Ghods}, {Ringuette}, {Robnett}, {Rosie}, {Santos}, {Sims},
  {Smith}, {Syce}, {Tegmark}, {Thyagarajan}, {Williams}, \&
  {Zheng}}]{dillon/etal:2020}
{Dillon}, J.~S., {Lee}, M., {Ali}, Z.~S., {et~al.} 2020, \mnras, 499, 5840,
  \dodoi{10.1093/mnras/staa3001}

\bibitem[{{Eastwood} {et~al.}(2018){Eastwood}, {Anderson}, {Monroe},
  {Hallinan}, {Barsdell}, {Bourke}, {Clark}, {Ellingson}, {Dowell}, {Garsden},
  {Greenhill}, {Hartman}, {Kocz}, {Lazio}, {Price}, {Schinzel}, {Taylor},
  {Vedantham}, {Wang}, \& {Woody}}]{eastwood/etal:2018}
{Eastwood}, M.~W., {Anderson}, M.~M., {Monroe}, R.~M., {et~al.} 2018, \aj, 156,
  32, \dodoi{10.3847/1538-3881/aac721}

\bibitem[{{Fagnoni} {et~al.}(2021){Fagnoni}, {de Lera Acedo}, {DeBoer},
  {Abdurashidova}, {Aguirre}, {Alexander}, {Ali}, {Balfour}, {Beardsley},
  {Bernardi}, {Billings}, {Bowman}, {Bradley}, {Bull}, {Burba}, {Carilli},
  {Cheng}, {Dexter}, {Dillon}, {Ewall-Wice}, {Fritz}, {Furlanetto},
  {Gale-Sides}, {Glendenning}, {Gorthi}, {Greig}, {Grobbelaar}, {Halday},
  {Hazelton}, {Hewitt}, {Hickish}, {Jacobs}, {Josaitis}, {Julius}, {Kern},
  {Kerrigan}, {Kim}, {Kittiwisit}, {Kohn}, {Kolopanis}, {Lanman}, {Plante},
  {Lekalake}, {Liu}, {MacMahon}, {Malan}, {Malgas}, {Maree}, {Martinot},
  {Matsetela}, {Mena Parra}, {Mesinger}, {Molewa}, {Morales}, {Mosiane},
  {Neben}, {Nikolic}, {Parsons}, {Patra}, {Pieterse}, {Pober}, {Razavi-Ghods},
  {Robnett}, {Rosie}, {Sims}, {Smith}, {Syce}, {Thyagarajan}, {Williams}, \&
  {Zheng}}]{fagnoni/etal:2021}
{Fagnoni}, N., {de Lera Acedo}, E., {DeBoer}, D.~R., {et~al.} 2021, \mnras,
  500, 1232, \dodoi{10.1093/mnras/staa3268}

\bibitem[{{G{\'o}rski} {et~al.}(2005){G{\'o}rski}, {Hivon}, {Banday},
  {Wandelt}, {Hansen}, {Reinecke}, \& {Bartelmann}}]{gorski/etal:2005}
{G{\'o}rski}, K.~M., {Hivon}, E., {Banday}, A.~J., {et~al.} 2005, \apj, 622,
  759, \dodoi{10.1086/427976}

\bibitem[{{Haslam} {et~al.}(1982){Haslam}, {Salter}, {Stoffel}, \&
  {Wilson}}]{haslam/etal:1982}
{Haslam}, C.~G.~T., {Salter}, C.~J., {Stoffel}, H., \& {Wilson}, W.~E. 1982,
  \aaps, 47, 1

\bibitem[{{Hazelton} {et~al.}(2017){Hazelton}, {Jacobs}, {Pober}, \&
  {Beardsley}}]{hazelton/etal:2017}
{Hazelton}, B.~J., {Jacobs}, D.~C., {Pober}, J.~C., \& {Beardsley}, A.~P. 2017,
  The Journal of Open Source Software, 2, 140, \dodoi{10.21105/joss.00140}

\bibitem[{{Hinshaw} {et~al.}(2013){Hinshaw}, {Larson}, {Komatsu}, {Spergel},
  {Bennett}, {Dunkley}, {Nolta}, {Halpern}, {Hill}, {Odegard}, {Page}, {Smith},
  {Weiland}, {Gold}, {Jarosik}, {Kogut}, {Limon}, {Meyer}, {Tucker}, {Wollack},
  \& {Wright}}]{hinshaw/etal:2013}
{Hinshaw}, G., {Larson}, D., {Komatsu}, E., {et~al.} 2013, \apjs, 208, 19,
  \dodoi{10.1088/0067-0049/208/2/19}

\bibitem[{{H{\"o}gbom}(1974)}]{hogbom:1974}
{H{\"o}gbom}, J.~A. 1974, \aaps, 15, 417

\bibitem[{{Hurley-Walker} {et~al.}(2017){Hurley-Walker}, {Callingham},
  {Hancock}, {Franzen}, {Hindson}, {Kapi{\'n}ska}, {Morgan}, {Offringa},
  {Wayth}, {Wu}, {Zheng}, {Murphy}, {Bell}, {Dwarakanath}, {For}, {Gaensler},
  {Johnston-Hollitt}, {Lenc}, {Procopio}, {Staveley-Smith}, {Ekers}, {Bowman},
  {Briggs}, {Cappallo}, {Deshpande}, {Greenhill}, {Hazelton}, {Kaplan},
  {Lonsdale}, {McWhirter}, {Mitchell}, {Morales}, {Morgan}, {Oberoi}, {Ord},
  {Prabu}, {Shankar}, {Srivani}, {Subrahmanyan}, {Tingay}, {Webster},
  {Williams}, \& {Williams}}]{hurley-walker/etal:2017}
{Hurley-Walker}, N., {Callingham}, J.~R., {Hancock}, P.~J., {et~al.} 2017,
  \mnras, 464, 1146, \dodoi{10.1093/mnras/stw2337}

\bibitem[{{Hurley-Walker} {et~al.}(2019){Hurley-Walker}, {Hancock}, {Franzen},
  {Callingham}, {Offringa}, {Hindson}, {Wu}, {Bell}, {For}, {Gaensler},
  {Johnston-Hollitt}, {Kapi{\'n}ska}, {Morgan}, {Murphy}, {McKinley},
  {Procopio}, {Staveley-Smith}, {Wayth}, \& {Zheng}}]{hurley-walker/etal:2019}
{Hurley-Walker}, N., {Hancock}, P.~J., {Franzen}, T.~M.~O., {et~al.} 2019,
  \pasa, 36, e047, \dodoi{10.1017/pasa.2019.37}

\bibitem[{{Kern} {et~al.}(2019){Kern}, {Parsons}, {Dillon}, {Lanman},
  {Fagnoni}, \& {de Lera Acedo}}]{kern/etal:2019}
{Kern}, N.~S., {Parsons}, A.~R., {Dillon}, J.~S., {et~al.} 2019, \apj, 884,
  105, \dodoi{10.3847/1538-4357/ab3e73}

\bibitem[{{Kern} {et~al.}(2020{\natexlab{a}}){Kern}, {Dillon}, {Parsons},
  {Carilli}, {Bernardi}, {Abdurashidova}, {Aguirre}, {Alexander}, {Ali},
  {Balfour}, {Beardsley}, {Billings}, {Bowman}, {Bradley}, {Bull}, {Burba},
  {Carey}, {Cheng}, {DeBoer}, {Dexter}, {de Lera Acedo}, {Ely}, {Ewall-Wice},
  {Fagnoni}, {Fritz}, {Furlanetto}, {Gale-Sides}, {Glendenning}, {Gorthi},
  {Greig}, {Grobbelaar}, {Halday}, {Hazelton}, {Hewitt}, {Hickish}, {Jacobs},
  {Julius}, {Kerrigan}, {Kittiwisit}, {Kohn}, {Kolopanis}, {Lanman}, {La
  Plante}, {Lekalake}, {Liu}, {MacMahon}, {Malan}, {Malgas}, {Maree},
  {Martinot}, {Matsetela}, {Mesinger}, {Molewa}, {Morales}, {Mosiane},
  {Murray}, {Neben}, {Nikolic}, {Nunhokee}, {Patra}, {Pieterse}, {Pober},
  {Razavi-Ghods}, {Ringuette}, {Robnett}, {Rosie}, {Sims}, {Smith}, {Syce},
  {Thyagarajan}, {Williams}, \& {Zheng}}]{kern/etal:2020}
{Kern}, N.~S., {Dillon}, J.~S., {Parsons}, A.~R., {et~al.} 2020{\natexlab{a}},
  \apj, 890, 122, \dodoi{10.3847/1538-4357/ab67bc}

\bibitem[{{Kern} {et~al.}(2020{\natexlab{b}}){Kern}, {Parsons}, {Dillon},
  {Lanman}, {Liu}, {Bull}, {Ewall-Wice}, {Abdurashidova}, {Aguirre},
  {Alexander}, {Ali}, {Balfour}, {Beardsley}, {Bernardi}, {Bowman}, {Bradley},
  {Burba}, {Carilli}, {Cheng}, {DeBoer}, {Dexter}, {de Lera Acedo}, {Fagnoni},
  {Fritz}, {Furlanetto}, {Glendenning}, {Gorthi}, {Greig}, {Grobbelaar},
  {Halday}, {Hazelton}, {Hewitt}, {Hickish}, {Jacobs}, {Julius}, {Kerrigan},
  {Kittiwisit}, {Kohn}, {Kolopanis}, {La Plante}, {Lekalake}, {MacMahon},
  {Malan}, {Malgas}, {Maree}, {Martinot}, {Matsetela}, {Mesinger}, {Molewa},
  {Morales}, {Mosiane}, {Murray}, {Neben}, {Parsons}, {Patra}, {Pieterse},
  {Pober}, {Razavi-Ghods}, {Ringuette}, {Robnett}, {Rosie}, {Sims}, {Smith},
  {Syce}, {Thyagarajan}, {Williams}, \& {Zheng}}]{kern/etal:2020b}
{Kern}, N.~S., {Parsons}, A.~R., {Dillon}, J.~S., {et~al.} 2020{\natexlab{b}},
  \apj, 888, 70, \dodoi{10.3847/1538-4357/ab5e8a}

\bibitem[{{Kolopanis} {et~al.}(2019){Kolopanis}, {Jacobs}, {Cheng}, {Parsons},
  {Kohn}, {Pober}, {Aguirre}, {Ali}, {Bernardi}, {Bradley}, {Carilli},
  {DeBoer}, {Dexter}, {Dillon}, {Kerrigan}, {Klima}, {Liu}, {MacMahon},
  {Moore}, {Thyagarajan}, {Nunhokee}, {Walbrugh}, \&
  {Walker}}]{kolopnis/etal:2019}
{Kolopanis}, M., {Jacobs}, D.~C., {Cheng}, C., {et~al.} 2019, \apj, 883, 133,
  \dodoi{10.3847/1538-4357/ab3e3a}

\bibitem[{{Lanman} {et~al.}(2019){Lanman}, {Hazelton}, {Jacobs}, {Kolopanis},
  {Pober}, {Aguirre}, \& {Thyagarajan}}]{lanman/etal:2019}
{Lanman}, A., {Hazelton}, B., {Jacobs}, D., {et~al.} 2019, The Journal of Open
  Source Software, 4, 1234, \dodoi{10.21105/joss.01234}

\bibitem[{{Li} {et~al.}(2019){Li}, {Pober}, {Barry}, {Hazelton}, {Morales},
  {Trott}, {Lanman}, {Wilensky}, {Sullivan}, {Beardsley}, {Booler}, {Bowman},
  {Byrne}, {Crosse}, {Emrich}, {Franzen}, {Hasegawa}, {Horsley},
  {Johnston-Hollitt}, {Jacobs}, {Jordan}, {Joseph}, {Kaneuji}, {Kaplan},
  {Kenney}, {Kubota}, {Line}, {Lynch}, {McKinley}, {Mitchell}, {Murray},
  {Pallot}, {Pindor}, {Rahimi}, {Riding}, {Sleap}, {Steele}, {Takahashi},
  {Tingay}, {Walker}, {Wayth}, {Webster}, {Williams}, {Wu}, {Wyithe},
  {Yoshiura}, \& {Zheng}}]{Li/etal:2019}
{Li}, W., {Pober}, J.~C., {Barry}, N., {et~al.} 2019, \apj, 887, 141,
  \dodoi{10.3847/1538-4357/ab55e4}

\bibitem[{{Liu} \& {Shaw}(2020)}]{liu/shaw:2020}
{Liu}, A., \& {Shaw}, J.~R. 2020, \pasp, 132, 062001,
  \dodoi{10.1088/1538-3873/ab5bfd}

\bibitem[{{Liu} {et~al.}(2016){Liu}, {Zhang}, \&
  {Parsons}}]{liu/zhang/parsons:2016}
{Liu}, A., {Zhang}, Y., \& {Parsons}, A.~R. 2016, \apj, 833, 242,
  \dodoi{10.3847/1538-4357/833/2/242}

\bibitem[{{Mesinger}(2019)}]{mesinger:2016}
{Mesinger}, A. 2019, {The Cosmic 21-cm Revolution; Charting the first billion
  years of our universe}, \dodoi{10.1088/2514-3433/ab4a73}

\bibitem[{{Morales} {et~al.}(2019){Morales}, {Beardsley}, {Pober}, {Barry},
  {Hazelton}, {Jacobs}, \& {Sullivan}}]{morales/etal:2019}
{Morales}, M.~F., {Beardsley}, A., {Pober}, J., {et~al.} 2019, \mnras, 483,
  2207, \dodoi{10.1093/mnras/sty2844}

\bibitem[{{Morales} \& {Hewitt}(2004)}]{morales/hewitt:2004}
{Morales}, M.~F., \& {Hewitt}, J. 2004, \apj, 615, 7, \dodoi{10.1086/424437}

\bibitem[{{Morales} \& {Matejek}(2009)}]{morales/matejek:2009}
{Morales}, M.~F., \& {Matejek}, M. 2009, \mnras, 400, 1814,
  \dodoi{10.1111/j.1365-2966.2009.15537.x}

\bibitem[{{Morales} \& {Wyithe}(2010)}]{morales/wyithe:2010}
{Morales}, M.~F., \& {Wyithe}, J. S.~B. 2010, \araa, 48, 127,
  \dodoi{10.1146/annurev-astro-081309-130936}

\bibitem[{{Mozdzen} {et~al.}(2017){Mozdzen}, {Bowman}, {Monsalve}, \&
  {Rogers}}]{mozdzen/etal:2017}
{Mozdzen}, T.~J., {Bowman}, J.~D., {Monsalve}, R.~A., \& {Rogers}, A.~E.~E.
  2017, \mnras, 464, 4995, \dodoi{10.1093/mnras/stw2696}

\bibitem[{{Newburgh} {et~al.}(2016){Newburgh}, {Bandura}, {Bucher}, {Chang},
  {Chiang}, {Cliche}, {Dav{\'e}}, {Dobbs}, {Clarkson}, {Ganga}, {Gogo},
  {Gumba}, {Gupta}, {Hilton}, {Johnstone}, {Karastergiou}, {Kunz}, {Lokhorst},
  {Maartens}, {Macpherson}, {Mdlalose}, {Moodley}, {Ngwenya}, {Parra},
  {Peterson}, {Recnik}, {Saliwanchik}, {Santos}, {Sievers}, {Smirnov},
  {Stronkhorst}, {Taylor}, {Vanderlinde}, {Van Vuuren}, {Weltman}, \&
  {Witzemann}}]{newburgh/etal:2016}
{Newburgh}, L.~B., {Bandura}, K., {Bucher}, M.~A., {et~al.} 2016, in Society of
  Photo-Optical Instrumentation Engineers (SPIE) Conference Series, Vol. 9906,
  Ground-based and Airborne Telescopes VI, ed. H.~J. {Hall}, R.~{Gilmozzi}, \&
  H.~K. {Marshall}, 99065X, \dodoi{10.1117/12.2234286}

\bibitem[{{Parsons} {et~al.}(2012){Parsons}, {Pober}, {Aguirre}, {Carilli},
  {Jacobs}, \& {Moore}}]{parsons/etal:2012}
{Parsons}, A.~R., {Pober}, J.~C., {Aguirre}, J.~E., {et~al.} 2012, \apj, 756,
  165, \dodoi{10.1088/0004-637X/756/2/165}

\bibitem[{{Parsons} {et~al.}(2010){Parsons}, {Backer}, {Foster}, {Wright},
  {Bradley}, {Gugliucci}, {Parashare}, {Benoit}, {Aguirre}, {Jacobs},
  {Carilli}, {Herne}, {Lynch}, {Manley}, \& {Werthimer}}]{parsons/etal:2010}
{Parsons}, A.~R., {Backer}, D.~C., {Foster}, G.~S., {et~al.} 2010, \aj, 139,
  1468, \dodoi{10.1088/0004-6256/139/4/1468}

\bibitem[{{Patil} {et~al.}(2017){Patil}, {Yatawatta}, {Koopmans}, {de Bruyn},
  {Brentjens}, {Zaroubi}, {Asad}, {Hatef}, {Jeli{\'c}}, {Mevius}, {Offringa},
  {Pandey}, {Vedantham}, {Abdalla}, {Brouw}, {Chapman}, {Ciardi}, {Gehlot},
  {Ghosh}, {Harker}, {Iliev}, {Kakiichi}, {Majumdar}, {Mellema}, {Silva},
  {Schaye}, {Vrbanec}, \& {Wijnholds}}]{patil/etal:2017}
{Patil}, A.~H., {Yatawatta}, S., {Koopmans}, L.~V.~E., {et~al.} 2017, \apj,
  838, 65, \dodoi{10.3847/1538-4357/aa63e7}

\bibitem[{{Planck Collaboration} {et~al.}(2020){Planck Collaboration},
  {Aghanim}, {Akrami}, {Arroja}, {Ashdown}, {Aumont}, {Baccigalupi},
  {Ballardini}, {Banday}, {Barreiro}, {Bartolo}, {Basak}, {Battye}, {Benabed},
  {Bernard}, {Bersanelli}, {Bielewicz}, {Bock}, {Bond}, {Borrill}, {Bouchet},
  {Boulanger}, {Bucher}, {Burigana}, {Butler}, {Calabrese}, {Cardoso},
  {Carron}, {Casaponsa}, {Challinor}, {Chiang}, {Colombo}, {Combet},
  {Contreras}, {Crill}, {Cuttaia}, {de Bernardis}, {de Zotti}, {Delabrouille},
  {Delouis}, {D{\'e}sert}, {Di Valentino}, {Dickinson}, {Diego}, {Donzelli},
  {Dor{\'e}}, {Douspis}, {Ducout}, {Dupac}, {Efstathiou}, {Elsner},
  {En{\ss}lin}, {Eriksen}, {Falgarone}, {Fantaye}, {Fergusson},
  {Fernandez-Cobos}, {Finelli}, {Forastieri}, {Frailis}, {Franceschi},
  {Frolov}, {Galeotta}, {Galli}, {Ganga}, {G{\'e}nova-Santos}, {Gerbino},
  {Ghosh}, {Gonz{\'a}lez-Nuevo}, {G{\'o}rski}, {Gratton}, {Gruppuso},
  {Gudmundsson}, {Hamann}, {Handley}, {Hansen}, {Helou}, {Herranz},
  {Hildebrandt}, {Hivon}, {Huang}, {Jaffe}, {Jones}, {Karakci}, {Keih{\"a}nen},
  {Keskitalo}, {Kiiveri}, {Kim}, {Kisner}, {Knox}, {Krachmalnicoff}, {Kunz},
  {Kurki-Suonio}, {Lagache}, {Lamarre}, {Langer}, {Lasenby}, {Lattanzi},
  {Lawrence}, {Le Jeune}, {Leahy}, {Lesgourgues}, {Levrier}, {Lewis},
  {Liguori}, {Lilje}, {Lilley}, {Lindholm}, {L{\'o}pez-Caniego}, {Lubin}, {Ma},
  {Mac{\'\i}as-P{\'e}rez}, {Maggio}, {Maino}, {Mandolesi}, {Mangilli},
  {Marcos-Caballero}, {Maris}, {Martin}, {Martinelli},
  {Mart{\'\i}nez-Gonz{\'a}lez}, {Matarrese}, {Mauri}, {McEwen}, {Meerburg},
  {Meinhold}, {Melchiorri}, {Mennella}, {Migliaccio}, {Millea}, {Mitra},
  {Miville-Desch{\^e}nes}, {Molinari}, {Moneti}, {Montier}, {Morgante}, {Moss},
  {Mottet}, {M{\"u}nchmeyer}, {Natoli}, {N{\o}rgaard-Nielsen}, {Oxborrow},
  {Pagano}, {Paoletti}, {Partridge}, {Patanchon}, {Pearson}, {Peel}, {Peiris},
  {Perrotta}, {Pettorino}, {Piacentini}, {Polastri}, {Polenta}, {Puget},
  {Rachen}, {Reinecke}, {Remazeilles}, {Renault}, {Renzi}, {Rocha}, {Rosset},
  {Roudier}, {Rubi{\~n}o-Mart{\'\i}n}, {Ruiz-Granados}, {Salvati}, {Sandri},
  {Savelainen}, {Scott}, {Shellard}, {Shiraishi}, {Sirignano}, {Sirri},
  {Spencer}, {Sunyaev}, {Suur-Uski}, {Tauber}, {Tavagnacco}, {Tenti},
  {Terenzi}, {Toffolatti}, {Tomasi}, {Trombetti}, {Valiviita}, {Van Tent},
  {Vibert}, {Vielva}, {Villa}, {Vittorio}, {Wandelt}, {Wehus}, {White},
  {White}, {Zacchei}, \& {Zonca}}]{planck/etal:2020}
{Planck Collaboration}, {Aghanim}, N., {Akrami}, Y., {et~al.} 2020, \aap, 641,
  A1, \dodoi{10.1051/0004-6361/201833880}

\bibitem[{{Pober} {et~al.}(2016){Pober}, {Hazelton}, {Beardsley}, {Barry},
  {Martinot}, {Sullivan}, {Morales}, {Bell}, {Bernardi}, {Bhat}, {Bowman},
  {Briggs}, {Cappallo}, {Carroll}, {Corey}, {de Oliveira-Costa}, {Deshpande},
  {Dillon}, {Emrich}, {Ewall-Wice}, {Feng}, {Goeke}, {Greenhill}, {Hewitt},
  {Hindson}, {Hurley-Walker}, {Jacobs}, {Johnston-Hollitt}, {Kaplan}, {Kasper},
  {Kim}, {Kittiwisit}, {Kratzenberg}, {Kudryavtseva}, {Lenc}, {Line}, {Loeb},
  {Lonsdale}, {Lynch}, {McKinley}, {McWhirter}, {Mitchell}, {Morgan}, {Neben},
  {Oberoi}, {Offringa}, {Ord}, {Paul}, {Pindor}, {Prabu}, {Procopio}, {Riding},
  {Rogers}, {Roshi}, {Sethi}, {Udaya Shankar}, {Srivani}, {Subrahmanyan},
  {Tegmark}, {Thyagarajan}, {Tingay}, {Trott}, {Waterson}, {Wayth}, {Webster},
  {Whitney}, {Williams}, {Williams}, \& {Wyithe}}]{pober/etal:2016}
{Pober}, J.~C., {Hazelton}, B.~J., {Beardsley}, A.~P., {et~al.} 2016, \apj,
  819, 8, \dodoi{10.3847/0004-637X/819/1/8}

\bibitem[{{Pritchard} \& {Loeb}(2012)}]{pritchard/loeb:2012}
{Pritchard}, J.~R., \& {Loeb}, A. 2012, Reports on Progress in Physics, 75,
  086901, \dodoi{10.1088/0034-4885/75/8/086901}

\bibitem[{{Rahimi} {et~al.}(2021){Rahimi}, {Pindor}, {Line}, {Barry}, {Trott},
  {Webster}, {Jordan}, {Wilensky}, {Yoshiura}, {Beardsley}, {Bowman}, {Byrne},
  {Chokshi}, {Hazelton}, {Hasegawa}, {Howard}, {Greig}, {Jacobs}, {Joseph},
  {Kolopanis}, {Lynch}, {McKinley}, {Mitchell}, {Murray}, {Morales}, {Pober},
  {Takahashi}, {Tingay}, {Wayth}, {Wyithe}, \& {Zheng}}]{rahimi/etal:2021}
{Rahimi}, M., {Pindor}, B., {Line}, J.~L.~B., {et~al.} 2021, \mnras, 508, 5954,
  \dodoi{10.1093/mnras/stab2918}

\bibitem[{{Rau} \& {Cornwell}(2011)}]{rau/cornwell:2011}
{Rau}, U., \& {Cornwell}, T.~J. 2011, \aap, 532, A71,
  \dodoi{10.1051/0004-6361/201117104}

\bibitem[{{Remazeilles} {et~al.}(2015){Remazeilles}, {Dickinson}, {Banday},
  {Bigot-Sazy}, \& {Ghosh}}]{remazeilles/etal:2015}
{Remazeilles}, M., {Dickinson}, C., {Banday}, A.~J., {Bigot-Sazy}, M.~A., \&
  {Ghosh}, T. 2015, \mnras, 451, 4311, \dodoi{10.1093/mnras/stv1274}

\bibitem[{{Shaw} {et~al.}(2014){Shaw}, {Sigurdson}, {Pen}, {Stebbins}, \&
  {Sitwell}}]{shaw/etal:2014}
{Shaw}, J.~R., {Sigurdson}, K., {Pen}, U.-L., {Stebbins}, A., \& {Sitwell}, M.
  2014, \apj, 781, 57, \dodoi{10.1088/0004-637X/781/2/57}

\bibitem[{{Sullivan} {et~al.}(2012){Sullivan}, {Morales}, {Hazelton}, {Arcus},
  {Barnes}, {Bernardi}, {Briggs}, {Bowman}, {Bunton}, {Cappallo}, {Corey},
  {Deshpande}, {deSouza}, {Emrich}, {Gaensler}, {Goeke}, {Greenhill}, {Herne},
  {Hewitt}, {Johnston-Hollitt}, {Kaplan}, {Kasper}, {Kincaid}, {Koenig},
  {Kratzenberg}, {Lonsdale}, {Lynch}, {McWhirter}, {Mitchell}, {Morgan},
  {Oberoi}, {Ord}, {Pathikulangara}, {Prabu}, {Remillard}, {Rogers}, {Roshi},
  {Salah}, {Sault}, {Udaya Shankar}, {Srivani}, {Stevens}, {Subrahmanyan},
  {Tingay}, {Wayth}, {Waterson}, {Webster}, {Whitney}, {Williams}, {Williams},
  \& {Wyithe}}]{sullivan/etal:2012}
{Sullivan}, I.~S., {Morales}, M.~F., {Hazelton}, B.~J., {et~al.} 2012, \apj,
  759, 17, \dodoi{10.1088/0004-637X/759/1/17}

\bibitem[{{Tegmark}(1997)}]{tegmark/maps:1997}
{Tegmark}, M. 1997, \apjl, 480, L87, \dodoi{10.1086/310631}

\bibitem[{{The HERA Collaboration} {et~al.}(2021){The HERA Collaboration},
  {Abdurashidova}, {Aguirre}, {Alexander}, {Ali}, {Balfour}, {Beardsley},
  {Bernardi}, {Billings}, {Bowman}, {Bradley}, {Bull}, {Burba}, {Carey},
  {Carilli}, {Cheng}, {DeBoer}, {Dexter}, {de Lera Acedo}, {Dibblee-Barkman},
  {Dillon}, {Ely}, {Ewall-Wice}, {Fagnoni}, {Fritz}, {Furlanetto},
  {Gale-Sides}, {Glendenning}, {Gorthi}, {Greig}, {Grobbelaar}, {Halday},
  {Hazelton}, {Hewitt}, {Hickish}, {Jacobs}, {Julius}, {Kern}, {Kerrigan},
  {Kittiwisit}, {Kohn}, {Kolopanis}, {Lanman}, {La Plante}, {Lekalake},
  {Lewis}, {Liu}, {MacMahon}, {Malan}, {Malgas}, {Maree}, {Martinot},
  {Matsetela}, {Mesinger}, {Molewa}, {Morales}, {Mosiane}, {Murray}, {Neben},
  {Nikolic}, {Nunhokee}, {Parsons}, {Patra}, {Pascua}, {Pieterse}, {Pober},
  {Razavi-Ghods}, {Ringuette}, {Robnett}, {Rosie}, {Sims}, {Singh}, {Smith},
  {Syce}, {Thyagarajan}, {Williams}, \& {Zheng}}]{hera/etal:2021}
{The HERA Collaboration}, {Abdurashidova}, Z., {Aguirre}, J.~E., {et~al.} 2021,
  arXiv e-prints, arXiv:2108.02263.
\newblock \doarXiv{2108.02263}

\bibitem[{{Thompson} {et~al.}(2017){Thompson}, {Moran}, \&
  {Swenson}}]{thompson/moran/swenson:2017}
{Thompson}, A.~R., {Moran}, J.~M., \& {Swenson}, George~W., J. 2017,
  {Interferometry and Synthesis in Radio Astronomy, 3rd Edition} (Springer,
  Cham), \dodoi{10.1007/978-3-319-44431-4}

\bibitem[{{Tingay} {et~al.}(2013){Tingay}, {Goeke}, {Bowman}, {Emrich}, {Ord},
  {Mitchell}, {Morales}, {Booler}, {Crosse}, {Wayth}, {Lonsdale}, {Tremblay},
  {Pallot}, {Colegate}, {Wicenec}, {Kudryavtseva}, {Arcus}, {Barnes},
  {Bernardi}, {Briggs}, {Burns}, {Bunton}, {Cappallo}, {Corey}, {Deshpande},
  {Desouza}, {Gaensler}, {Greenhill}, {Hall}, {Hazelton}, {Herne}, {Hewitt},
  {Johnston-Hollitt}, {Kaplan}, {Kasper}, {Kincaid}, {Koenig}, {Kratzenberg},
  {Lynch}, {Mckinley}, {Mcwhirter}, {Morgan}, {Oberoi}, {Pathikulangara},
  {Prabu}, {Remillard}, {Rogers}, {Roshi}, {Salah}, {Sault}, {Udaya-Shankar},
  {Schlagenhaufer}, {Srivani}, {Stevens}, {Subrahmanyan}, {Waterson},
  {Webster}, {Whitney}, {Williams}, {Williams}, \& {Wyithe}}]{tingay/etal:2013}
{Tingay}, S.~J., {Goeke}, R., {Bowman}, J.~D., {et~al.} 2013, \pasa, 30, e007,
  \dodoi{10.1017/pasa.2012.007}

\bibitem[{{Trott} {et~al.}(2016){Trott}, {Pindor}, {Procopio}, {Wayth},
  {Mitchell}, {McKinley}, {Tingay}, {Barry}, {Beardsley}, {Bernardi}, {Bowman},
  {Briggs}, {Cappallo}, {Carroll}, {de Oliveira-Costa}, {Dillon}, {Ewall-Wice},
  {Feng}, {Greenhill}, {Hazelton}, {Hewitt}, {Hurley-Walker},
  {Johnston-Hollitt}, {Jacobs}, {Kaplan}, {Kim}, {Lenc}, {Line}, {Loeb},
  {Lonsdale}, {Morales}, {Morgan}, {Neben}, {Thyagarajan}, {Oberoi},
  {Offringa}, {Ord}, {Paul}, {Pober}, {Prabu}, {Riding}, {Udaya Shankar},
  {Sethi}, {Srivani}, {Subrahmanyan}, {Sullivan}, {Tegmark}, {Webster},
  {Williams}, {Williams}, {Wu}, \& {Wyithe}}]{trott/etal:2016}
{Trott}, C.~M., {Pindor}, B., {Procopio}, P., {et~al.} 2016, \apj, 818, 139,
  \dodoi{10.3847/0004-637X/818/2/139}

\bibitem[{{van Haarlem} {et~al.}(2013){van Haarlem}, {Wise}, {Gunst}, {Heald},
  {McKean}, {Hessels}, {de Bruyn}, {Nijboer}, {Swinbank}, {Fallows},
  {Brentjens}, {Nelles}, {Beck}, {Falcke}, {Fender}, {H{\"o}randel},
  {Koopmans}, {Mann}, {Miley}, {R{\"o}ttgering}, {Stappers}, {Wijers},
  {Zaroubi}, {van den Akker}, {Alexov}, {Anderson}, {Anderson}, {van Ardenne},
  {Arts}, {Asgekar}, {Avruch}, {Batejat}, {B{\"a}hren}, {Bell}, {Bell}, {van
  Bemmel}, {Bennema}, {Bentum}, {Bernardi}, {Best}, {B{\^\i}rzan}, {Bonafede},
  {Boonstra}, {Braun}, {Bregman}, {Breitling}, {van de Brink}, {Broderick},
  {Broekema}, {Brouw}, {Br{\"u}ggen}, {Butcher}, {van Cappellen}, {Ciardi},
  {Coenen}, {Conway}, {Coolen}, {Corstanje}, {Damstra}, {Davies}, {Deller},
  {Dettmar}, {van Diepen}, {Dijkstra}, {Donker}, {Doorduin}, {Dromer}, {Drost},
  {van Duin}, {Eisl{\"o}ffel}, {van Enst}, {Ferrari}, {Frieswijk}, {Gankema},
  {Garrett}, {de Gasperin}, {Gerbers}, {de Geus}, {Grie{\ss}meier}, {Grit},
  {Gruppen}, {Hamaker}, {Hassall}, {Hoeft}, {Holties}, {Horneffer}, {van der
  Horst}, {van Houwelingen}, {Huijgen}, {Iacobelli}, {Intema}, {Jackson},
  {Jelic}, {de Jong}, {Juette}, {Kant}, {Karastergiou}, {Koers}, {Kollen},
  {Kondratiev}, {Kooistra}, {Koopman}, {Koster}, {Kuniyoshi}, {Kramer},
  {Kuper}, {Lambropoulos}, {Law}, {van Leeuwen}, {Lemaitre}, {Loose}, {Maat},
  {Macario}, {Markoff}, {Masters}, {McFadden}, {McKay-Bukowski}, {Meijering},
  {Meulman}, {Mevius}, {Middelberg}, {Millenaar}, {Miller-Jones}, {Mohan},
  {Mol}, {Morawietz}, {Morganti}, {Mulcahy}, {Mulder}, {Munk}, {Nieuwenhuis},
  {van Nieuwpoort}, {Noordam}, {Norden}, {Noutsos}, {Offringa}, {Olofsson},
  {Omar}, {Orr{\'u}}, {Overeem}, {Paas}, {Pandey-Pommier}, {Pandey}, {Pizzo},
  {Polatidis}, {Rafferty}, {Rawlings}, {Reich}, {de Reijer}, {Reitsma},
  {Renting}, {Riemers}, {Rol}, {Romein}, {Roosjen}, {Ruiter}, {Scaife}, {van
  der Schaaf}, {Scheers}, {Schellart}, {Schoenmakers}, {Schoonderbeek},
  {Serylak}, {Shulevski}, {Sluman}, {Smirnov}, {Sobey}, {Spreeuw}, {Steinmetz},
  {Sterks}, {Stiepel}, {Stuurwold}, {Tagger}, {Tang}, {Tasse}, {Thomas},
  {Thoudam}, {Toribio}, {van der Tol}, {Usov}, {van Veelen}, {van der Veen},
  {ter Veen}, {Verbiest}, {Vermeulen}, {Vermaas}, {Vocks}, {Vogt}, {de Vos},
  {van der Wal}, {van Weeren}, {Weggemans}, {Weltevrede}, {White}, {Wijnholds},
  {Wilhelmsson}, {Wucknitz}, {Yatawatta}, {Zarka}, {Zensus}, \& {van
  Zwieten}}]{vanhaarlem/etal:2013}
{van Haarlem}, M.~P., {Wise}, M.~W., {Gunst}, A.~W., {et~al.} 2013, \aap, 556,
  A2, \dodoi{10.1051/0004-6361/201220873}

\bibitem[{{Wayth} {et~al.}(2015){Wayth}, {Lenc}, {Bell}, {Callingham},
  {Dwarakanath}, {Franzen}, {For}, {Gaensler}, {Hancock}, {Hindson},
  {Hurley-Walker}, {Jackson}, {Johnston-Hollitt}, {Kapi{\'n}ska}, {McKinley},
  {Morgan}, {Offringa}, {Procopio}, {Staveley-Smith}, {Wu}, {Zheng}, {Trott},
  {Bernardi}, {Bowman}, {Briggs}, {Cappallo}, {Corey}, {Deshpande}, {Emrich},
  {Goeke}, {Greenhill}, {Hazelton}, {Kaplan}, {Kasper}, {Kratzenberg},
  {Lonsdale}, {Lynch}, {McWhirter}, {Mitchell}, {Morales}, {Morgan}, {Oberoi},
  {Ord}, {Prabu}, {Rogers}, {Roshi}, {Shankar}, {Srivani}, {Subrahmanyan},
  {Tingay}, {Waterson}, {Webster}, {Whitney}, {Williams}, \&
  {Williams}}]{wayth/etal:2015}
{Wayth}, R.~B., {Lenc}, E., {Bell}, M.~E., {et~al.} 2015, \pasa, 32, e025,
  \dodoi{10.1017/pasa.2015.26}

\bibitem[{{Ye} {et~al.}(2021){Ye}, {Gull}, {Tan}, \& {Nikolic}}]{ye/etal:2021}
{Ye}, H., {Gull}, S.~F., {Tan}, S.~M., \& {Nikolic}, B. 2021, arXiv e-prints,
  arXiv:2101.11172.
\newblock \doarXiv{2101.11172}

\bibitem[{{Zheng} {et~al.}(2017{\natexlab{a}}){Zheng}, {Tegmark}, {Dillon},
  {Liu}, {Neben}, {Tribiano}, {Bradley}, {Buza}, {Ewall-Wice}, {Gharibyan},
  {Hickish}, {Kunz}, {Losh}, {Lutomirski}, {Morgan}, {Narayanan}, {Perko},
  {Rosner}, {Sanchez}, {Schutz}, {Valdez}, {Villasenor}, {Yang}, {Zarb Adami},
  {Zelko}, \& {Zheng}}]{zheng/etal:2017b}
{Zheng}, H., {Tegmark}, M., {Dillon}, J.~S., {et~al.} 2017{\natexlab{a}},
  \mnras, 465, 2901, \dodoi{10.1093/mnras/stw2910}

\bibitem[{{Zheng} {et~al.}(2017{\natexlab{b}}){Zheng}, {Tegmark}, {Dillon},
  {Kim}, {Liu}, {Neben}, {Jonas}, {Reich}, \& {Reich}}]{zheng/etal:2017}
---. 2017{\natexlab{b}}, \mnras, 464, 3486, \dodoi{10.1093/mnras/stw2525}

\end{thebibliography}
\bibliographystyle{aasjournal}


\end{CJK*}
\end{document}